\documentclass{emulateapj}
\input{epsf}
\submitted{The Astrophysical Journal, accepted}

\shorttitle{The stellar structure and kinematics of dwarf spheroidals}

\shortauthors{{\L}okas et al.}

\begin{document}

\title{The stellar structure and kinematics of dwarf spheroidal galaxies\\ formed by tidal stirring}

\author{Ewa L. {\L}okas\altaffilmark{1}, Stelios Kazantzidis\altaffilmark{2,3,4}, Jaros{\l}aw Klimentowski\altaffilmark{1},
\\ Lucio Mayer\altaffilmark{5,6}, and Simone Callegari\altaffilmark{5}}

\altaffiltext{1}{Nicolaus Copernicus Astronomical Center, Warsaw, Poland; lokas@camk.edu.pl}
\altaffiltext{2}{Center for Cosmology and Astro-Particle Physics,
    The Ohio State University, Columbus, OH 43210, USA; stelios@mps.ohio-state.edu}
\altaffiltext{3}{Department of Physics,
    The Ohio State University, Columbus, OH 43210, USA}
\altaffiltext{4}{Department of Astronomy,
    The Ohio State University, Columbus, OH 43210, USA}
\altaffiltext{5}{Institute for Theoretical Physics, University of Z\"urich, CH-8057 Z\"urich, Switzerland;
        lucio@phys.ethz.ch}
\altaffiltext{6}{Institute of Astronomy, Department of Physics, ETH Z\"urich, CH-8093 Z\"urich, Switzerland}

\begin{abstract}

Using high-resolution $N$-body simulations we study the stellar properties of dwarf spheroidal (dSph)
galaxies resulting from the tidally induced morphological transformation of disky dwarfs on a cosmologically motivated
eccentric orbit around the Milky Way. The dwarf galaxy models initially consist of an exponential stellar
disk embedded in an extended spherical dark matter halo. Depending on the initial orientation of the disk with respect
to the orbital plane, different final configurations are obtained. The least evolved dwarf is triaxial and retains a
significant amount of rotation. The more evolved dwarfs are prolate spheroids with little rotation. We
show that in this scenario the final density distribution of stars can be approximated by a simple modification
of the Plummer law. The kinematics of the dwarfs is significantly different depending on the line of sight which has
important implications for mapping the observed stellar velocity dispersions of dwarfs to subhalo circular velocities.
When the dwarfs are observed along the long axis, the measured velocity dispersion is higher and decreases faster with
radius. In the case where rotation is significant, when viewed perpendicular to the long axis, the effect of minor axis
rotation is detected, as expected for triaxial systems. We model the velocity dispersion profiles and rotation curves
of the dwarfs under the assumption of constant mass-to-light ratio by solving the Jeans equations for spherical and
axisymmetric systems and adjusting different sets of free parameters, including the total mass. We find that the mass
is typically overestimated when the dwarf is seen along the long axis and underestimated when the observation is along
the short or intermediate axis. For the studied cases the effect of non-sphericity cannot however bias the inferred
mass by more than 60 percent in either direction, even for the most strongly stripped dwarf which is close to
disruption.
\end{abstract}


\section{Introduction}

Dwarf spheroidal (dSph) galaxies of the Local Group (for a review see Mateo 1998) can provide crucial tests of the
currently favored cold dark matter (CDM) paradigm of cosmological structure formation (e.g., White \& Rees 1978;
Blumenthal et al. 1984). Several scenarios have been proposed so far for the origin of these systems, including
gravitational interactions in the form of galaxy harassment (e.g., Moore et al. 1996) tidal stirring (Mayer et al.
2001; Kravtsov et al. 2004; Kazantzidis et al. 2004b) or resonant stripping (D'Onghia et al. 2009) as well as
hydrodynamical processes (e.g., Ricotti \& Gnedin 2005; Mayer et al. 2007; Tassis et al. 2008). The tidal stirring
scenario proposed by Mayer et al. (2001, see also Gnedin et al. 1999; Pe\~narrubia  et al. 2008) can explain the
morphological transformation of disky dwarf galaxies resembling dwarf irregulars into pressure-supported stellar
systems with the structural properties of dSphs. When combined with ram pressure stripping and the effect of the UV
background (Bullock et al. 2000; Susa \& Umemura 2004) it can also account for the low gas content and the very high
mass-to-light ratios ($M/L$) of some of them (Mayer et al. 2006; 2007).

In order to explore this scenario in more detail Klimentowski et al. (2007, 2009a) have
recently studied the evolution of two-component dwarf galaxies orbiting the Milky Way on a cosmologically
motivated eccentric orbit using high resolution $N$-body simulations. The dwarfs were initially composed of a stellar
disk embedded in a more extended dark matter halo. Three simulations were performed which differed by the initial
inclination of the disk with respect to the orbital plane. Although the evolution proceeded a little differently in each
case, strong mass loss and morphological transformation of the disk always took place. After the first or second (out
of five) pericenter passages the disk evolves into a bar which in the subsequent evolution becomes shorter. The
morphological transformation of the stellar component is accompanied by the loss of angular momentum so that little
rotation remains at the end. The final products are very similar in shape to the dSph galaxies of the Local Group. In
this paper we study in detail the structural and kinematic properties of stars in these dwarfs at the end of their
tidal evolution.

\begin{table*}
\caption{Properties of the simulated dwarfs.}
\label{simprop}
\begin{center}
\begin{tabular}{cccccccc}
\hline
\hline
simulation & $i$[deg]& $r_{\rm max}$[kpc] & $M_{\rm tot}(r_{\rm max}) [$M$_\odot]$ & $N_{\rm stars} (r_{\rm max})$ &
$N_{\rm DM} (r_{\rm max})$ & $M_{\rm stars}(r_{\rm max}) [$M$_\odot]$ & $M_{\rm DM}(r_{\rm max}) [$M$_\odot]$ \\
\hline
A & 90 & 2.00 & $3.73 \times 10^7$ & 73423 & 25394 & $1.10 \times 10^7$ & $2.63 \times 10^7$ \\
B & 45 & 1.70 & $2.05 \times 10^7$ & 37739 & 14378 & $5.64 \times 10^6$ & $1.49 \times 10^7$ \\
C & 0  & 1.44 & $4.67 \times 10^6$ & 8379  & 3298  & $1.25 \times 10^6$ & $3.42 \times 10^6$ \\
\hline
\end{tabular}
\end{center}
\end{table*}

DSph galaxies are also interesting because of their significant dark matter content. It turns out that the tidal
stirring scenario is able to reproduce both the mildly dark matter dominated dwarfs like Fornax or Leo I (Klimentowski
et al. 2007, 2009a) and the very dark ones like Draco (Mayer et al. 2007). In spite of significant mass loss due to
tidal stripping the dwarfs are able to retain enough dark matter to remain mostly dark, although the extended dark halo
is lost very fast and the mass approximately traces light in the final stages. This means that according to this
scenario the mass modeling of dSph galaxies can be reliably done adopting the scale lengths of the distribution of
light as is indeed a common approach.

The presence of tidal stripping however also means that in addition to the contamination from Milky Way stars, which
can be dealt with using photometric methods (Majewski et al. 2000) or metallicities (Walker et al. 2009a), the kinematic
samples used for dynamical modeling can be contaminated by tidal debris. Klimentowski et al. (2009b) have shown that
such contamination is indeed very probable because of the typically radial orientation of tidal tails for dwarfs on
eccentric orbits. This contamination may artificially inflate the velocity dispersion and bias the inferred parameters
of the density profile or the anisotropy of stellar orbits. It can however be reliably removed using procedures for
interloper rejection (Klimentowski et al. 2007; {\L}okas et al. 2008).

Another source of systematic error in the dynamical modeling may be due to departures from sphericity. As mentioned
above, the products of tidal stirring are not spherical but rather form prolate spheroids. In this work we study the
effect of non-sphericity on mass and anisotropy estimates, i.e. we show what biases are expected if spherical models
are applied to galaxies whose shapes depart from spherical. We also discuss how the estimates can improve if
axisymmetric models are used in those cases where rotation is detected.

The paper is organized as follows. In section 2 we provide a brief description of the simulations used in this work.
In section 3 we study the properties of the stellar component of three dSph galaxy models obtained in three simulations
with different initial inclination of the disk. Section 4 discusses the line-of-sight density and velocity
distributions, as they would be seen by a distant observer. In section 5 we briefly compare the properties of our
simulated dwarfs to observations. Section 6 is devoted to modeling the mock kinematic data using spherical and
axisymmetric models and studying the biases inherent in such modeling for different lines of sight. The discussion
follows in section 7.

\section{The Simulations}

\begin{figure}
\begin{center}
    \leavevmode
    \epsfxsize=8.5cm
    \epsfbox[0 35 285 450]{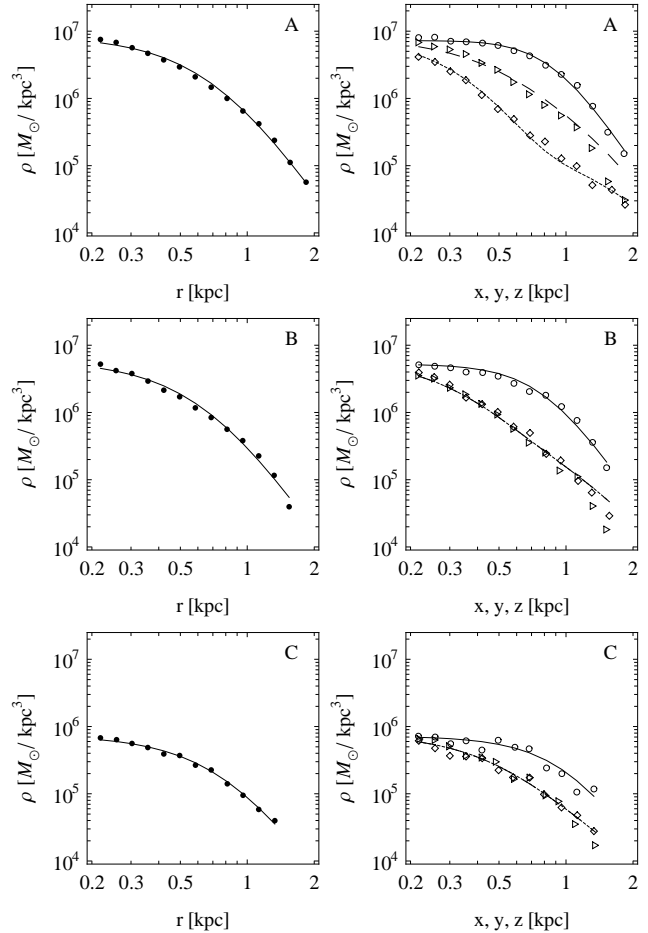}
\end{center}
\caption{Density distributions of stars in the simulated dwarfs. The first, second and third row
corresponds respectively to dwarf A, B and C of Table~\ref{simprop}.
The left column panels show the density distribution
of stars measured in radial bins (filled circles) and the solid lines show the Plummer law (\ref{plummerden})
with parameters from Table~\ref{fittedspherical} fitted to the data.
The second column presents the densities measured in narrow cuboids
along the $x$ (circles), $y$ (triangles) and $z$ (squares) axis of the dwarfs. The solid, dashed and
dotted lines show respectively the best-fitting profiles
obtained with density following from the potential (\ref{modplum}) with parameters
listed in Table~\ref{fittednonspherical}.}
\label{density}
\end{figure}

The simulations used for the present study follow the evolution of dwarf disk-like systems orbiting within the static
host potential of a Milky Way-sized galaxy and are described in detail in Klimentowski et al. (2007, 2009a). Here we
provide a short summary of the most important parameters. Live dwarf galaxy models are constructed using the technique
by Hernquist (1993) and consist of an exponential stellar disk embedded in a spherical and isotropic Navarro et al.
(1996, hereafter NFW) dark matter halo. The structural properties of the dark halo and disk are related through
disk galaxy formation models in the currently favored concordance CDM cosmological model (Mo et al. 1998).

The dwarf progenitor has the total mass $M = 4.3 \times 10^9$ M$_{\odot}$. The mass and radial scale length of the disk
are $1.5 \times 10^8$ M$_{\odot}$ and $R_{\rm d} = 1.3$ kpc, respectively, and its vertical structure is modeled by
isothermal sheets (with the vertical scale height $z_{\rm d} = 0.13$ kpc). The halo virial mass and concentration are
$M_{\rm vir} = 3.7 \times 10^9$ M$_{\odot}$ and $c=15$, respectively. The halo is exponentially truncated outside the
virial radius $r_{\rm vir} \simeq 40.2$ kpc to keep the total mass finite (Kazantzidis et al. 2004a) and adiabatically
contracted in response to the growth of the disk (Blumenthal et al. 1986). The host galaxy is modeled by a static
gravitational potential based on the dynamical mass model A1 for the Milky Way (Klypin et al. 2002). It consists
of a NFW halo with the virial mass of $M_{\rm vir} = 10^{12}$ M$_{\odot}$ and concentration $c=12$, a stellar disk with
mass $M_{\rm D} = 4 \times 10^{10}$ M$_{\odot}$, the scale length $R_{\rm d} = 3.5$ kpc and the scale height $z_{\rm d}
= 0.35$ kpc, and a bulge of mass $M_{\rm b}=0.008 M_{\rm vir}$ and scale-length $a_{\rm b} = 0.2 R_{\rm d}$. The dwarf
galaxy evolves on an eccentric orbit with apocenter $r_{\rm a}=120$ kpc and pericenter to apocenter ratio of $r_{\rm
p}/r_{\rm a} \approx 0.2$, close to the median ratio of pericentric to apocentric radii found in high-resolution
cosmological $N$-body simulations (Ghigna et al. 1998; Diemand et al. 2007). The evolution is followed for 10 Gyr
corresponding to approximately five orbital times. We used three simulations with the disk initially inclined by
0$^\circ$, 45$^\circ$, and 90$^\circ$ with respect to the orbital plane.

Lastly, the simulations were performed using PKDGRAV, a multistepping, parallel, tree $N$-body code (Stadel 2001). We
sampled the live dwarf galaxy with $4 \times 10^{6}$ dark matter particles and $10^{6}$ stellar particles, and employed
a gravitational softening length of $100$ and $50$ pc, respectively.

\begin{table}
\caption{Fitted parameters of the spherical models for the distribution of stars.}
\label{fittedspherical}
\begin{center}
\begin{tabular}{ccc}
\hline
\hline
simulation & $M_{\rm s} [$M$_\odot]$ & $a$ [kpc]  \\
\hline
A &  $1.33 \times 10^7$ & 0.73  \\
B &  $7.01 \times 10^6$ & 0.66  \\
C &  $2.00 \times 10^6$ & 0.86  \\
\hline
\end{tabular}
\end{center}
\end{table}

\begin{figure}
\begin{center}
    \leavevmode
    \epsfxsize=4.3cm
    \epsfbox[2 2 167 167]{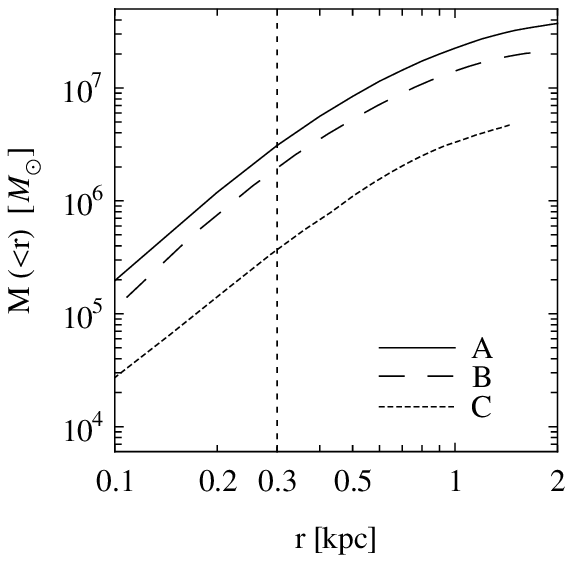}
    \leavevmode
    \epsfxsize=4.2cm
    \epsfbox[0 0 165 165]{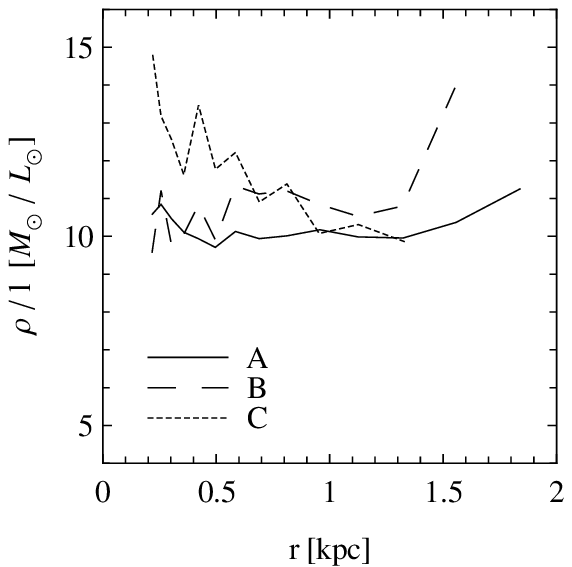}
\end{center}
\caption{Left panel: the final mass profiles of the three simulated dwarfs A, B and C (solid, dashed and dotted line
respectively).
The vertical dashed line indicates the values of mass within $300$ pc, $M_{300}$. Right panel: the final mass-to-light
density ratios
for the three dwarfs.
\label{mml}}
\end{figure}

\section{The products of tidal stirring}

Table~\ref{simprop} lists the basic properties of the simulated dwarfs in the final stage.
For a detailed description of the intermediate evolution we refer the reader to Klimentowski et al. (2009a).
In order to avoid any contamination from the tidally disrupted stars we select only the stellar and dark matter
particles within radius $r<r_{\rm max}$ kpc from the center of the dwarf. The adopted values of $r_{\rm max}$, listed in
the Table, were determined by inspection of the stellar density profiles, i.e. they are the radii at which the profiles
start to flatten signifying the transition to the tidal tails. The radii are of the order of the dwarf's tidal
radius or slightly smaller.  The Table also gives the initial inclinations of the
stellar disk plane with respect to the orbital plane, the total masses of the dwarfs within $r_{\rm max}$, the numbers
of stars and dark particles contained inside these radii and the stellar and dark masses separately. In section 5 we
discuss the observational properties of our dwarfs.

The left column of Figure~\ref{density} shows with the filled circles the radially averaged density profile of
the stars in the three simulated dwarfs measured in the range between $r=0.2$ kpc (which is larger than our resolution
limit) and $r=r_{\rm max}$ in logarithmic bins. The data are well fitted by the Plummer distribution
\begin{equation}
\label{plummerden}
        \rho(r) = \frac{3 a^2 M_{\rm s}}{4 \pi} \frac{1}{(r^2 + a^2)^{5/2}},
\end{equation}
where $M_{\rm s}$ is the total mass of the stellar component and $a$ is the characteristic scale-length.
This density follows via the Poisson equation from the potential
\begin{equation}        \label{plummerpot}
        \Phi(r) = -\frac{G M_{\rm s}}{(r^2 + a^2)^{1/2}}.
\end{equation}
The fits were done by minimizing the sum
$\Sigma_i (\rho_{\rm data} - \rho_{\rm model})^2/\rho_{\rm model}^2$. The best-fitting profiles are plotted as solid
lines in the left-column panels of Figure~\ref{density} and the best-fitting parameters are
listed in Table~\ref{fittedspherical}. The corresponding half-light radii are $r_{1/2}=1, 0.9$ and 1.15 kpc respectively
for model A, B and C.

In order to describe the departures from sphericity in the structure of the dwarfs
using the stellar particles we determine the principal axes of the inertia tensor and define our coordinate system so
that the $x$ axis lies along the longest axis of the stellar distribution and the $z$ axis along the shortest. We
measure the density along the $x$, $y$ and $z$ axis in cuboids of logarithmic bins along the given axis and of $\pm
0.2$ kpc width along the remaining dimensions, so for example for the measurements along the $z$ axis we adopt
$|x|<0.2$ kpc and $|y|<0.2$ kpc. The measurements were made along the positive and negative side of a given axis and
then averaged.

\begin{table}
\caption{Fitted parameters of the non-spherical models for the distribution of stars.}
\label{fittednonspherical}
\begin{center}
\begin{tabular}{cccccc}
\hline
\hline
simulation & $M_{\rm s} [$M$_\odot]$ & $a$ [kpc] & $b$ [kpc] & $c$ [kpc] & $d$ [kpc] \\
\hline
A &  $1.62 \times 10^7$ & 0.83 &  0.93 & 0.81 & 1.41 \\
B &  $8.74 \times 10^6$ & 0.76 &  0.74 & 1.16 & 1.16 \\
C &  $2.70 \times 10^6$ & 1.02 &  0.77 & 1.22 & 1.22 \\
\hline
\end{tabular}
\end{center}
\end{table}

\begin{figure*}
\begin{center}
    \leavevmode
    \epsfxsize=12.5cm
    \epsfbox[0 5 435 430]{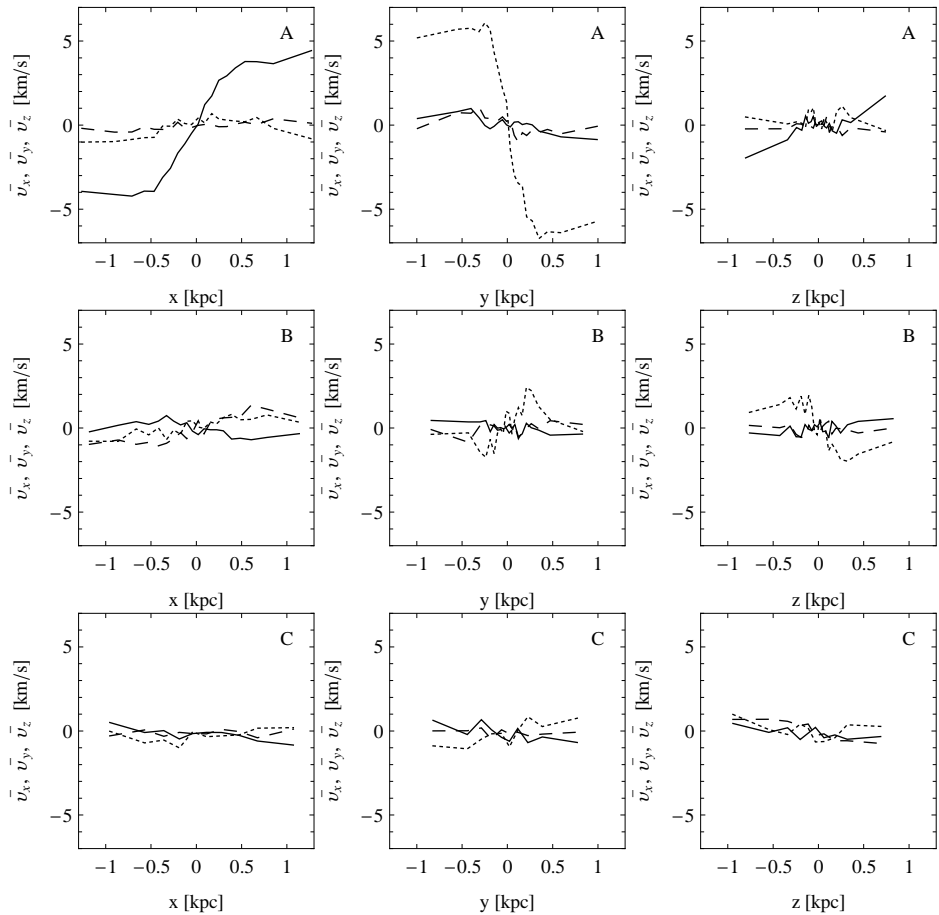}
\end{center}
\caption{Mean velocities along the three axes $\overline{v_x}$ (dotted line), $\overline{v_y}$ (solid line)
and $\overline{v_z}$ (dashed line)
as a function of $x$, $y$ and $z$ (left to right column). The three rows show results for dwarf A, B and C
respectively.}
\label{rotation}
\end{figure*}

\begin{figure*}
\begin{center}
    \leavevmode
    \epsfxsize=12.5cm
    \epsfbox[5 0 430 435]{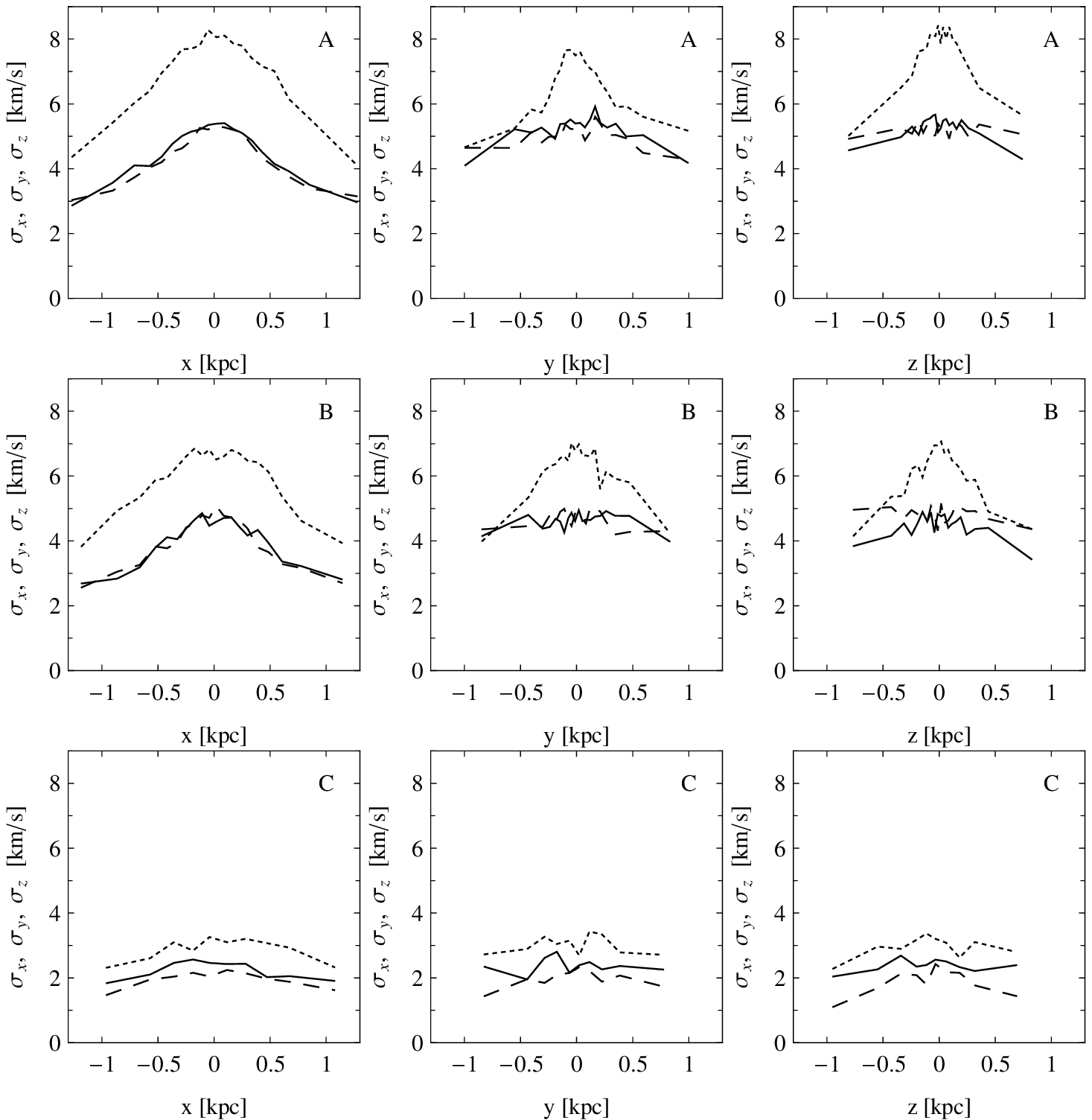}
\end{center}
\caption{Dispersions of the velocities along the three axes $\sigma_x$ (dotted line), $\sigma_y$ (solid line)
and $\sigma_z$ (dashed line)
as a function of $x$, $y$ and $z$ (left to right column). The three rows show results for dwarf A, B and C
respectively.}
\label{dispersion}
\end{figure*}

The density profiles measured along the three axes are shown in the right column of Figure~\ref{density} with circles
corresponding to the measurements along the $x$ axis, triangles along the $y$ axis and squares along $z$. We see that
the shape of model A is clearly triaxial: the densities are significantly different along the three axes: the
distribution is rather flat along the $x$ axis signifying the presence of a remnant bar and very steep (almost a
power law) along the $z$ axis. As we demonstrate below, the $z$ axis almost coincides with the rotation axis of the
dwarf. This means that in addition to the presence of the remnant bar, dwarf A is also flattened due to rotation.

The shapes of the models B and C are spheroidal, i.e. the density distribution is rather flat along the $x$ axis and
steeper along $y$ and $z$ but the latter two profiles are indistinguishable (therefore our choice of $y$ and $z$ is
random with respect to $x$). Note also that the distribution of stars in model C is noisier, the dwarf is
irregular and close to disruption, as indicated by its much smaller mass (see Table~\ref{simprop}).

The density distribution along the three axes can be well approximated by the density following by
the Poisson equation from the modified Plummer potential
\begin{eqnarray}
        &&\Phi(x,y,z) = \label{modplum} \\
        && -\frac{G M_{\rm s}}{[(x^2 + y^2 + z^2 + a^2)^2 - b^2 x^2 + c^2 y^2 + d^2 z^2]^{1/4}}  \nonumber
\end{eqnarray}
where again $M_{\rm s}$ is the total mass while $a, b, c$ and $d$ are constants with the dimension of length.
The first term in the denominator is the same as in the original Plummer potential, $(r^2 + a^2)$. The second,
negative term, $- b^2 x^2$, introduces a bar-like distribution along the $x$ axis and the third and fourth positive
terms $c^2 y^2$ and $d^2 z^2$, make the profile steeper along the $y$ and $z$ axis respectively. The form of the
potential (\ref{modplum}) is a generalization of the modified Plummer distribution proposed by Lynden-Bell (1962).

We fitted the profiles measured from the simulation with the density following from
(\ref{modplum}) via the Poisson equation by the same minimization scheme, except that now the
data points for model A were weighted by the density measured along the $y$ axis which have intermediate values
(otherwise the measurements along the $z$ axis would be fitted best). For models B and C the data points were weighted
by the average value of density along $x$ and $y$. The solid, dashed and dotted lines in the second column
of Figure~\ref{density} show respectively the best-fitting profiles along the $x$, $y$ and $z$ axis respectively,
obtained with parameters listed in Table~\ref{fittednonspherical}. Note that for models B and C we kept $c=d$.

The important dynamical properties of the three simulated dwarfs are illustrated further in Figure~\ref{mml}.
This figure shows the final mass, $M(<r)$, and mass-to-light density, $\rho/l$, profiles plotted up to each
corresponding $r_{\rm max}$. It can immediately be seen that the mass profiles are very
different in spite of the fact that the three dwarfs had identical initial density distributions and orbits
and differed only by the initial inclination of the disk with respect to the orbital plane. Interestingly,
while our dwarfs started with identical masses within $300$ pc, $M_{300}$, they become substantially different
among the three final models being almost an order of magnitude larger in dwarf A compared to that of dwarf C.

Figure~\ref{mml} also shows that the final mass-to-light density ratios are not significantly different among the three
dwarfs and, in addition, they are all almost constant with radius. The former is related to the fact that the
initial mass-to-light distributions were identical and indicates that the dark matter and stars are lost in proportional
amounts in all cases. The latter is a consequence of two facts. First, the initial mass-to-light distributions were
constant within $\sim 2$ kpc or so (Klimentowski et al. 2007). Second, after the outer part of the dark matter halo,
which was initially much more extended than the stars, is stripped at the first pericenter passage, the stars and dark
matter are stripped from all radii. Note that a similar result is obtained in simulations including gas dynamics (Mayer
et al. 2007). The final mass-to-light profile could however be different if other initial conditions were assumed
(e.g. Pe\~narrubia et al. 2008).

The velocity distribution of the stellar components of the dwarfs is illustrated in Figures~\ref{rotation} and
\ref{dispersion}. In Figure~\ref{rotation} we plot the mean velocities along the three axes, $\overline{v_x}$,
$\overline{v_y}$ and $\overline{v_z}$ as a function of $x$, $y$ and $z$. As in the case of density, the measurements
were made in narrow cuboids placed along a given axis, e.g. for the measurements as a function of $z$ we selected stars
with $|x|<0.2$ kpc and $|y|<0.2$ kpc. For model A significant rotation can be seen, but mostly in the $xy$ plane, i.e.
much less rotation is seen along $z$. For model B rotation is very small and for model C negligible. This can be
understood as due to more effective removal of angular momentum for models where the stars initially had angular
momentum more aligned with the orbital angular momentum of the dwarf (e.g., Read et al. 2006).

Figure~\ref{dispersion} shows the velocity dispersion profiles along the three axes $\sigma_x$, $\sigma_y$
and $\sigma_z$ as a function of $x$, $y$ and $z$. Here and below the dispersions are measured with respect to the
mean velocity in a given bin and calculated using the unbiased estimator $s = [\sum_{i=1}^n (v_i -
\overline{v})^2/(n-1)]^{1/2}$ where $n$ is the number of stars in a bin. We can see that the velocity dispersions
along $x$ are always largest, while those along $y$ and $z$ are smaller and comparable to each other.

%\vspace{0.1in}

\section{The line-of-sight distributions}

\begin{figure}
\begin{center}
    \leavevmode
    \epsfxsize=8cm
    \epsfbox[0 10 285 420]{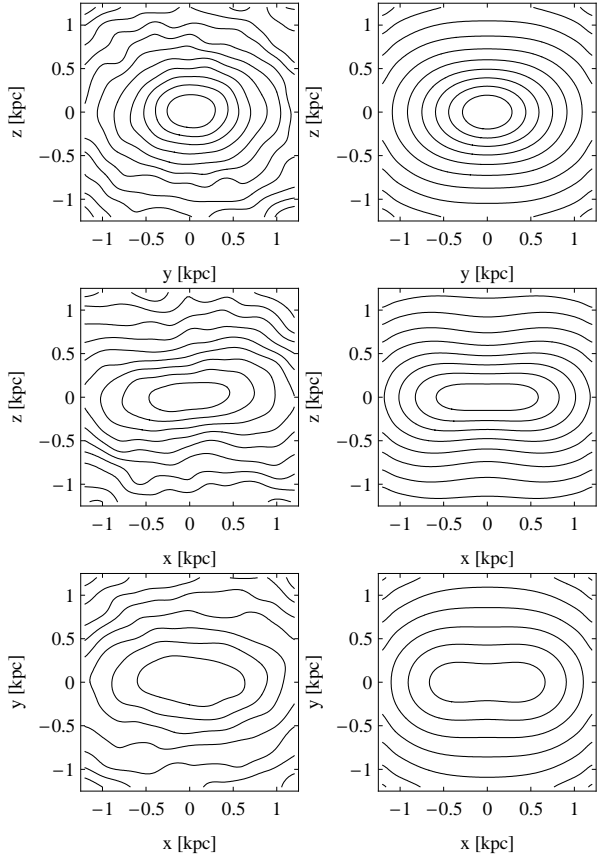}
\end{center}
\caption{Surface mass distribution of the stellar component of dwarf A observed along the $x$, $y$ and $z$ axis
(from the upper to the lower row). Left panels show the contours of equal surface density of stars
measured from the simulation. Right panels plot the corresponding distributions of the
modified Plummer model (\ref{modplum}) with parameters from Table~\ref{fittednonspherical}
adjusted to the density profiles
in Figure~\ref{density}. The surface densities $\Sigma$ were expressed in M$_{\odot}$ kpc$^{-2}$ and the contours
are spaced by $\Delta\log \Sigma =0.2$. The innermost contour level is $\log \Sigma =7$ (upper row), 6.8 (middle
row) and 6.6 (lower row).}
\label{surden90}
\end{figure}

In this section we discuss the line-of-sight distributions of density and velocity of the stars which can be directly
compared to observations. The triaxiality of dwarf A is further illustrated in Figure~\ref{surden90} where we plot the
surface density distributions of the stars seen along the three axes $x$, $y$ and $z$ from the top to the bottom row.
The left column panels show the isodensity contours of the surface density measured from the simulation and the right
column panels the corresponding values from the best-fitting model following from the modified Plummer potential
(\ref{modplum}) with parameters from Table~\ref{fittednonspherical}. We see that the model reproduces quite well the
basic features of the simulated dwarf. When viewed along the $z$ axis (lower row) the remnant bar is well
visible but there is otherwise little flattening. When viewed along the $y$ axis (middle row), both the bar and the
flattening along $z$ are well visible. When viewed along the $x$ axis (upper row) the bar is not seen, the contours are
almost circular except for some flattening along the $z$ axis.

\begin{figure}
\begin{center}
    \leavevmode
    \epsfxsize=8cm
    \epsfbox[0 5 280 280]{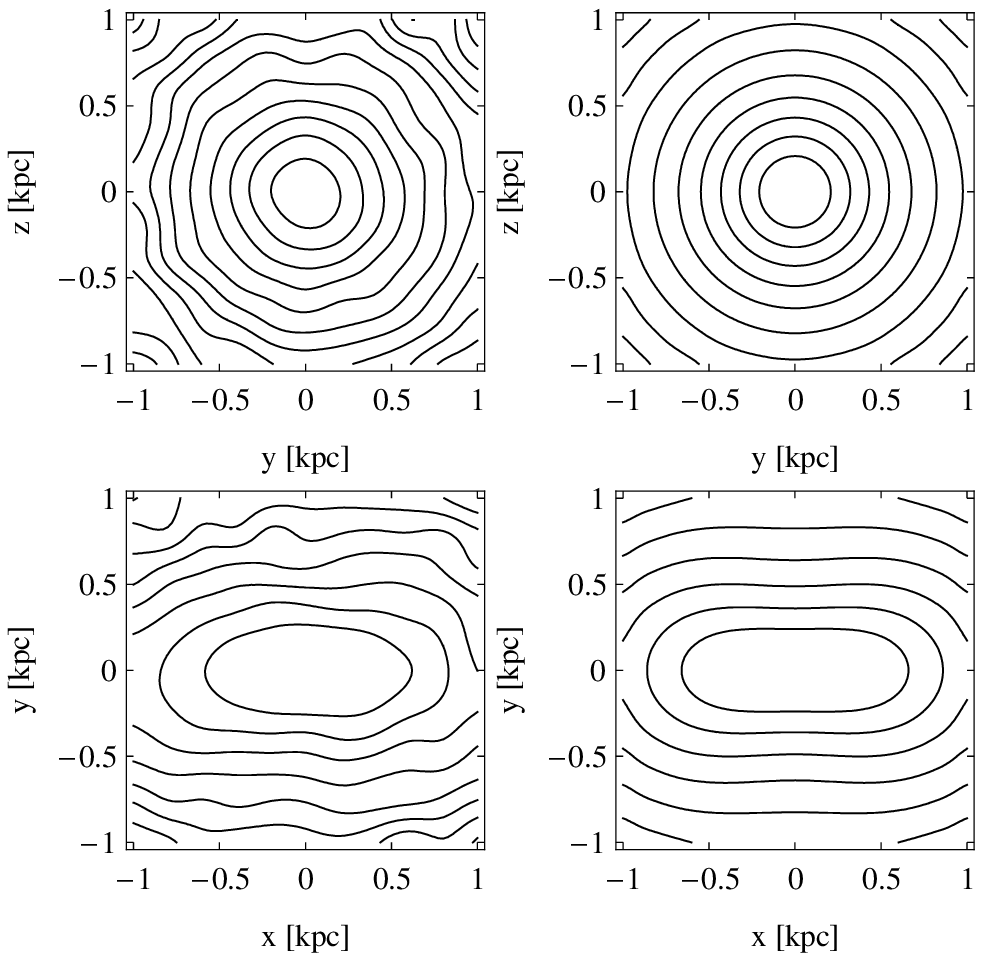}
\end{center}
\caption{Surface mass distribution of the stellar component of dwarf B observed along the $x$ axis
(upper row) and perpendicular to it (lower row). Left panels show the contours of equal surface density of stars
measured from the simulation. Right panels plot the corresponding distributions of the
modified Plummer model (\ref{modplum}) with parameters from Table~\ref{fittednonspherical}
adjusted to the density profiles
in Figure~\ref{density}. The surface densities $\Sigma$ were expressed in M$_{\odot}$ kpc$^{-2}$ and the contours
are spaced by $\Delta\log \Sigma =0.2$. The innermost contour level is $\log \Sigma =6.8$ (upper row) and
6.4 (lower row).}
\label{surden45}
\end{figure}

\begin{figure}
\begin{center}
    \leavevmode
    \epsfxsize=8cm
    \epsfbox[0 5 280 280]{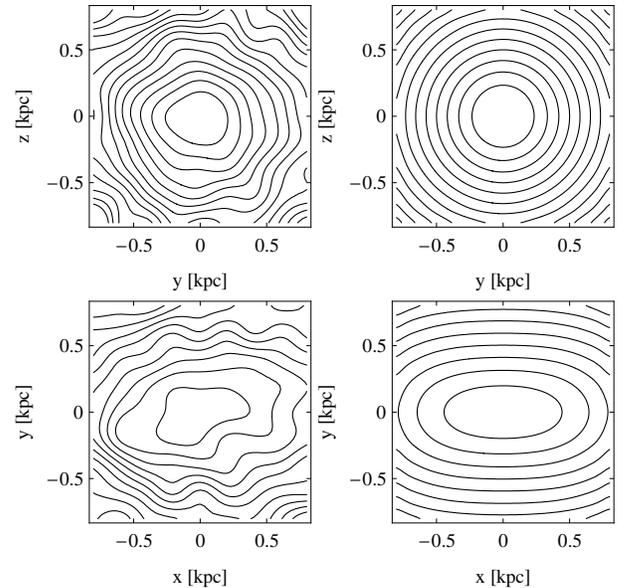}
\end{center}
\caption{The same as Figure~\ref{surden45} but for dwarf C. The contour spacing is $\Delta\log \Sigma =0.1$ and
the innermost contour level is $\log \Sigma =6$ (upper row) and 5.8 (lower row).}
\label{surden0}
\end{figure}

Figures~\ref{surden45} and \ref{surden0} present analogous results for models B and C, except that now we show only the
view along the longest axis (upper row) and along one of the other two axes (lower row) since the two look very
similar. As expected for spheroidal shapes, the surface density contours for the view along the $x$ axis are circular.
In the perpendicular direction an elongated shape is seen. The two models look quite similar, although model B is more
massive (see the different contour levels), except for the fact that in the case of model C the contours are much more
irregular, which is a sign of a stage close to disruption.

\begin{figure*}
\begin{center}
    \leavevmode
    \epsfxsize=12.5cm
    \epsfbox[20 60 550 830]{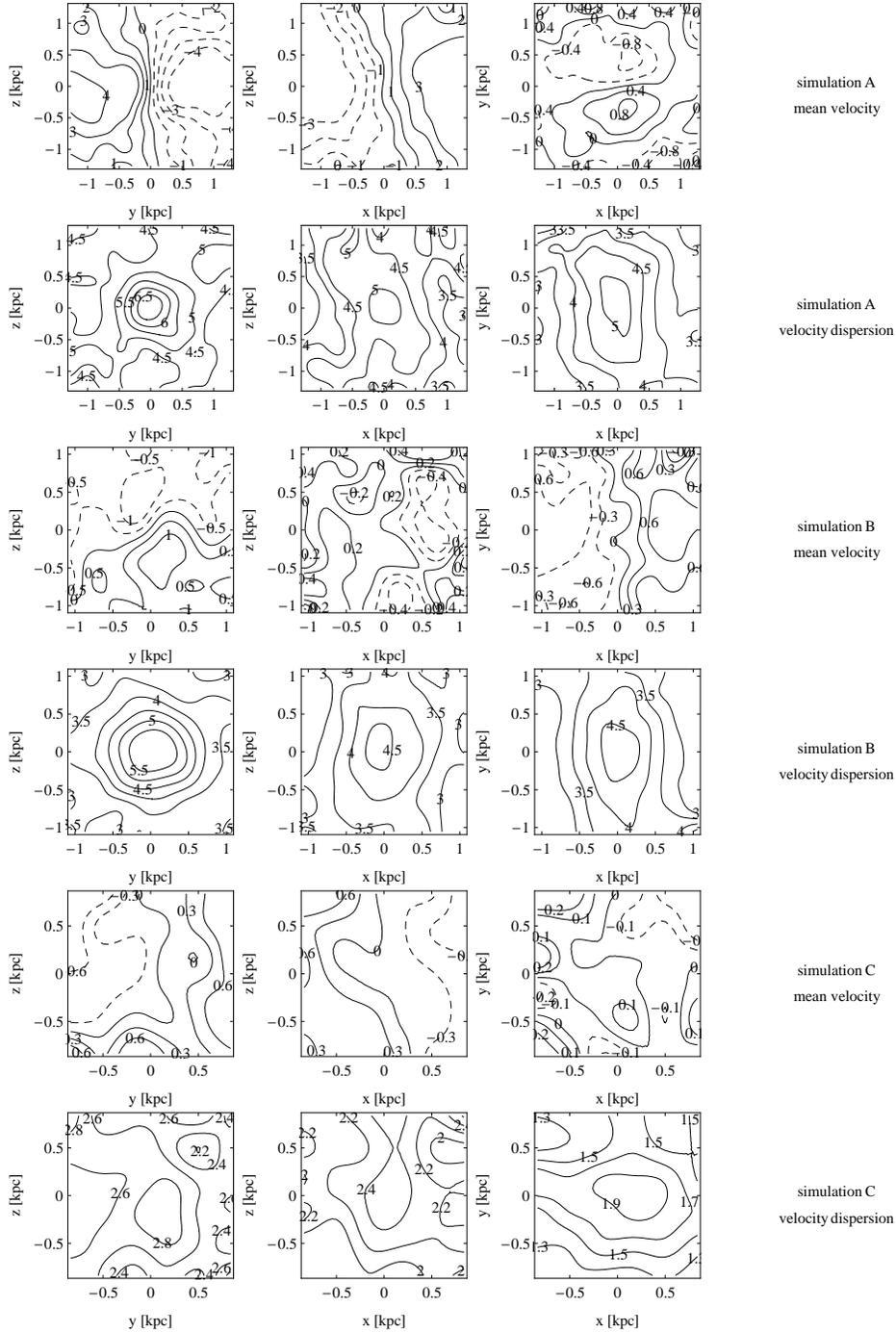}
\end{center}
\caption{Line-of-sight kinematics of the stellar component observed along the $x$, $y$ and $z$ axis
(from the left to the right column). The first and second row present the mean velocity and velocity dispersion
maps respectively, for the dwarf in simulation A. The subsequent rows show the results for simulations B and C as
marked on the right hand side of the Figure. Contour levels in km s$^{-1}$ are indicated in each panel. Contours
of negative mean velocity are shown with dashed lines. The measurements were made by binning the data
in cells of size 0.28, 0.3 and 0.33 kpc respectively for simulation A, B and C respectively.}
\label{velocity}
\end{figure*}

Figure~\ref{velocity} illustrates the line-of-sight kinematics of the stars in the simulated dwarfs. The measurements
are done in the same configuration as for the surface density in Figures~\ref{surden90}-\ref{surden0}. Note that no
clipping of the data is necessary since we still consider only stars within $r_{\rm max}$. The columns from the left to
the right correspond to observations along the $x$, $y$ and $z$ axis. The first row shows the contours of equal
line-of-sight velocity and the second row the contours of equal line-of-sight velocity dispersion for model A. Next
rows present analogous measurements for models B and C. Contour levels of the quantities expressed in km s$^{-1}$ are
marked in each panel and the dashed contour lines indicate negative values.

Again, for model A rotation is well visible
when the observation is along the $x$ axis (left column) and along $y$ axis (middle column) while it is much smaller
for the line of sight along the $z$ axis (right column). The velocity
dispersions are similar for the observations along $z$ and $y$ while the dispersion measured along the longest axis is
larger and decreases more steeply with distance from the center of the dwarf. A similar steeper decline of velocity
dispersion along the longest axis is seen for model B, but the rotation level is low in this case. For model C the
dispersion is rather flat even along the longest axis and rotation is very low.

\begin{table*}
\caption{Observational properties of the simulated dwarfs.}
\label{obsprop}
\begin{center}
\begin{tabular}{ccccccccc}
\hline
\hline
simulation & $M_V$ & $\mu_V [{\rm mag}/{\rm arcsec}^2]$ & $r_{1/2}$ [kpc] & $\sigma_0$ [km s$^{-1}$] & $V/\sigma$ \\
\hline
A &   -11.8 & 25.8-24.7 & 0.53-0.88 & 5-6.5   & 0.16-0.62  \\
B &   -11.1 & 26.0-25.2 & 0.47-0.74 & 4.5-5.5 & 0.09-0.18  \\
C & \  -9.7 & 27.9-27.4 & 0.73-0.98 & 1.9-2.8 & 0.11-0.25  \\
\hline
\end{tabular}
\end{center}
\end{table*}

The velocity structure seems to be the richest in the case of model A, as expected for triaxial systems.
Figures~\ref{rotation} and \ref{velocity} show that the orbital structure of our simulated dwarfs is similar to the one
found for simple triaxial models corresponding to St\"ackel potentials (de Zeeuw 1985; Statler 1987). For
such models four families of orbits are identified: box orbits, short-axis tube orbits and inner and outer long-axis
tube orbits. The only orbits with net angular momentum circulate either around the short or the long axis in agreement
with the rotation curves shown in Figure~\ref{rotation}. The total angular momentum is thus generally misaligned from
the short axis which results in a velocity gradient along the apparent minor axis when the galaxy is observed in
projection. This means that the zero velocity curve is misaligned with the minor axis, as we indeed see in the middle
panel of the upper row of Figure~\ref{velocity} (observation along $y$). The rotation is however much stronger around
the short axis. For prolate spheroids (models B and C) only the long-axis tubes are allowed but since the rotation
levels are low in these cases the stars must travel in both directions so that no single (clockwise or anti-clockwise)
direction dominates.

\vspace{0.1in}

\section{Comparison with observations}

To summarize the properties of the simulated dwarfs described in the two previous sections, in
Table~\ref{obsprop} we provide the basic parameters that can be directly compared to observations. The visual
magnitudes $M_V$ were calculated from the total stellar mass of the models assuming the stellar $M/L_V = 3$
M$_\odot$/L$_\odot$. For the remaining quantities, the central surface brightness $\mu_V$, the half-light radius
$r_{1/2}$, central velocity dispersion $\sigma_0$ and the ratio of the rotation velocity to the velocity dispersion
$V/\sigma$ we provide ranges of values corresponding to different lines of sight (along $x$, $y$ and $z$). The values
of $V/\sigma$ were calculated by taking the maximum line-of-sight rotation and dispersion measured along the major or
minor axis of the galaxy, as in Figure~\ref{velocity}.

Comparison with the observed values of these parameters for the Local Group dSph galaxies (e.g. Mateo 1998; Grebel et
al. 2003; Tolstoy et al. 2009) shows that they are rather representative for this type of objects. Only the
velocity dispersion values may appear a little too low. Note however, that we have considered here the final stages of
the evolution, i.e. after 10 Gyr when the dwarfs have lost majority of their mass. Although the progenitors of dSph
galaxies are expected to become satellites early, not all such galaxies have been tidally stripped for so long. Moving
back 2 or 4 Gyr in our simulations we would easily find earlier stages with velocity dispersions 50-100 percent higher
but other properties very similar (see section 4 and fig. 12 in Klimentowski et al. 2009a).

The parameters in Table~\ref{obsprop} also
obey the relationships like $M_V$-$\mu_V$ or $M_V$-$r_{1/2}$, i.e. in plots showing these relations (such as fig. 34 in
Kormendy et al. 2009 or fig. 1 in Tolstoy et al. 2009) our dwarfs would occupy regions characteristic of dSph galaxies.
Although with just three cases considered we cannot really explore correlations between the parameters, we can
immediately see that there is a clear trend of surface brightness decreasing with decreasing luminosity as observed in
real dSph galaxies, in contrast to ellipticals (fig. 1 in Kormendy et al. 2009). This feature, among others, has been
interpreted as pointing towards different formation scenarios of spheroidal and elliptical galaxies by Kormendy (1985)
and Kormendy et al. (2009). They suggested that elliptical galaxies may form mostly via mergers while spheroidals are
rather late-type systems that underwent transformation due to star-formation processes or environmental effects such as
tidal stirring. This idea has recently gained support from the theoretical side; based on the analysis of the simulated
Local Group Klimentowski et al. (2009c) found that mergers of subhaloes are quite rare in such systems and occur early
on so they cannot significantly contribute to the formation of a large fraction of dSph galaxies. The tidal
stirring scenario thus seems to be the most effective gravitational mechanism by which such objects could form.

Although our initial conditions constitute just one possible model for the progenitors of dSphs,
our results suggest that tidal interactions with the primary alone may introduce an appreciable scatter
in $M_{300}$ (see Figure~\ref{mml}), which could lead to some tension with claims of a common mass scale for Galactic
satellites within $300$ pc (Strigari et al. 2008). In addition, while we cannot exclude the effects of baryonic physics
on our findings, one could expect that a wider spectrum of initial conditions and orbits would likely further
differentiate the inner masses of dSph galaxies.

Recently, Walker et al. (2009c) suggested instead the existence of a scaling relation for dSph galaxies between the
half-light radius and the mass contained within it of the form $M(r_{1/2})/$M$_\odot = 5800 (r_{1/2}/{\rm pc})^{1.4}$.
The masses $M(r_{1/2})$ of our dwarfs (in units of $10^7$M$_\odot$) are within the ranges 0.94-2, 0.52-1 and 0.22-0.33
(depending on the line of sight) respectively for dwarfs A, B and C. Combining with the half-light radii from
Table~\ref{obsprop} we find that our dwarfs fall below the line proposed by Walker et al., i.e. for these half-light
radii they have smaller masses than the relation would predict. Note however that the masses used by Walker et al. may
be overestimated due to contamination. With just three cases considered here (including one close to disruption) we
cannot fully address the issue and we defer such investigations to future work.

\begin{figure*}
\begin{center}
    \leavevmode
    \epsfxsize=12.5cm
    \epsfbox[5 0 430 435]{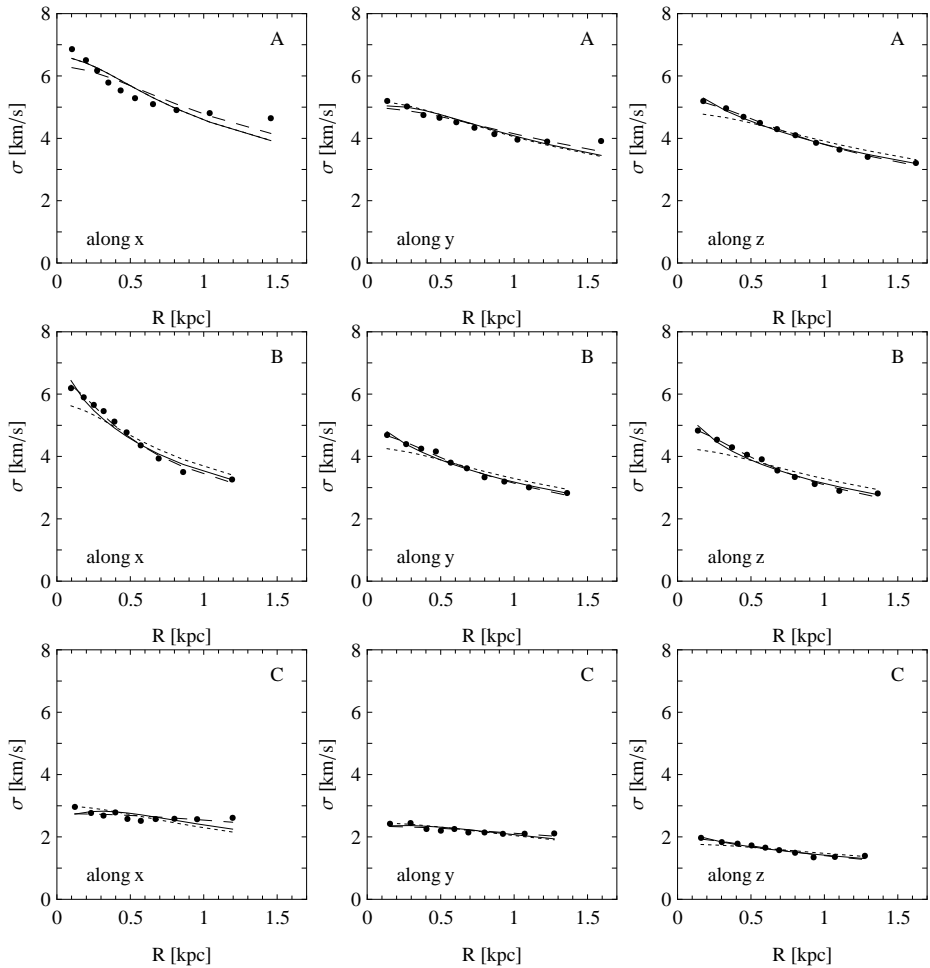}
\end{center}
\caption{Line-of-sight velocity dispersion profiles as a function of the projected radius $R$ for dwarfs obtained in
simulation A, B and C (from the upper to the lower row) for the observation
along the $x$, $y$ and $z$ axis (from the left to the right column). The dotted lines show the
best fitting isotropic ($\beta=0$) solutions of the Jeans equation for the one-component model with
scale length $a$ adopted from the fit of the surface density of stars (Table~\ref{fitted}
and Figure~\ref{surdenprof}) and only the mass $M$ adjusted. The solid lines are for models with $a$ assumed and
$M$ and anisotropy $\beta$ fitted. The dashed lines show the best-fitting isotropic models with $M$ and $a$ adjusted.
In the upper left panel the dotted line coincides with the solid one.
The best-fitting parameters are listed in Table~\ref{fittedsph}. }
\label{dispprof}
\end{figure*}

\section{The effect of non-sphericity on mass estimates}

\subsection {Velocity moments and the Jeans equations}

In order to further study the observational consequences of non-sphericity of dSph galaxies we now attempt to
model the kinematics of our simulated dwarfs. For each of the characteristic
lines of sight discussed above, i.e. along the $x, y$ and $z$ axis, we measure the rotation velocity $V$ and the
velocity dispersion $\sigma$ along the line of sight. For this purpose we use all the available stars although
in real observations the presently available samples are at least an order of magnitude less numerous. We are however
more interested here in studying systematic errors in the estimated parameters rather than estimating their
realistic statistical errors. We assign to the measured values of the velocity moments the true sampling errors
which are used to weigh the data points when fitting. Although these errors are unrealistically small in our
case, for real samples they would be relevant, i.e. they would be larger by a constant factor and data
points would be weighted by the same relative weights.

In the cases where the rotation level is very low we proceed in the standard
way: we bin the stars in ten equal-number bins along the projected radius $R$ (the distance from the center of the
dwarf to the star on the surface perpendicular to the line of sight). The velocity dispersion in each bin is calculated
using the standard unbiased estimator of dispersion and assigned a sampling error of size $\sigma/\sqrt{2(n-1)}$ where
$n$ is the number of stars per bin. In the cases when rotation is detected (observation along the $x$ and
$y$ axis for model A) we first transform the projected data using the symmetries of the axisymmetric system so that all
stars have positive coordinates. Then the data are again binned along the projected radius $R$. The measured velocity
dispersion profiles are plotted as dots as a function of the projected radius $R$ in Figure~\ref{dispprof}. The
sampling errors are not shown because they are very small.

\begin{figure}
\begin{center}
    \leavevmode
    \epsfxsize=8cm
    \epsfbox[0 0 300 300]{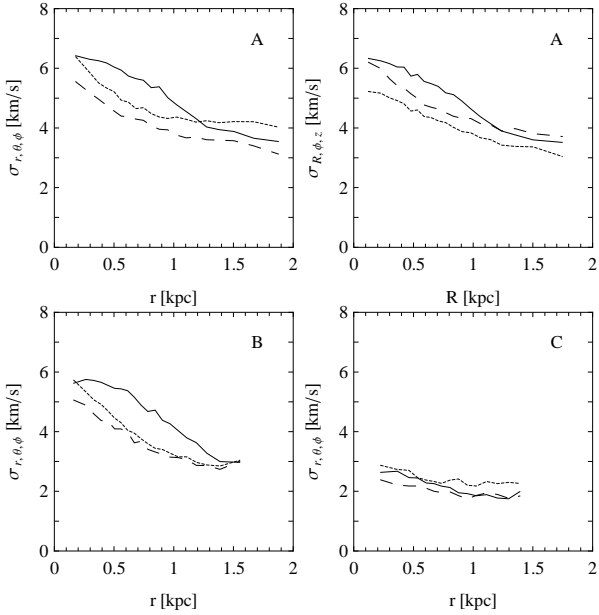}
\end{center}
\caption{The velocity dispersion profiles of the stellar component measured in spherical coordinates $r$, $\theta$,
$\phi$ in shells containing equal number of stars for models A (upper left panel), B (lower left panel) and C (lower
right panel). In each of these panels the solid line refers to $\sigma_r$, the dashed line to $\sigma_{\theta}$ and the
dotted line to $\sigma_{\phi}$. The upper right panel shows the velocity dispersion profiles of the stellar component
in dwarf A in cylindrical coordinates $R$, $\phi$ and $z$ as a function of $R$. The solid,
dashed and dotted line refers to $\sigma_R$, $\sigma_{\phi}$ and $\sigma_z$ respectively.}
\label{dispcomp}
\end{figure}

\begin{figure}
\begin{center}
    \leavevmode
    \epsfxsize=4.2cm
    \epsfbox[0 0 160 157]{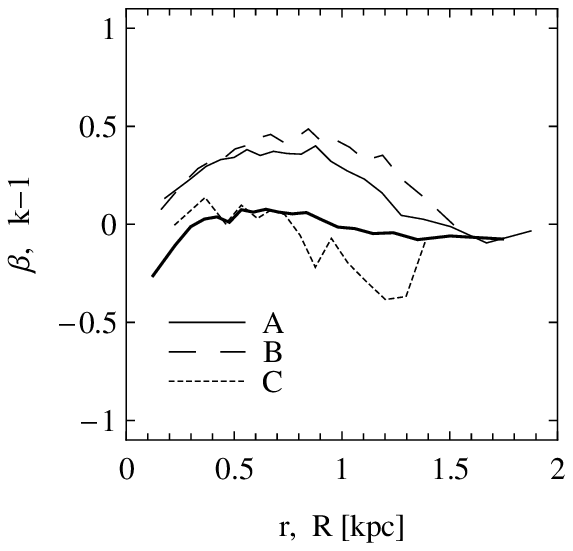}
    \leavevmode
    \epsfxsize=4.2cm
    \epsfbox[0 0 160 157]{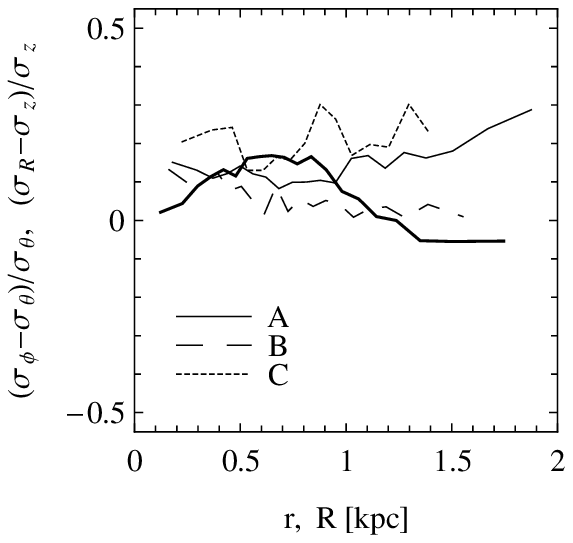}
\end{center}
\caption{Left panel: the spherical anisotropy profiles $\beta$ of the three simulated dwarfs A, B and C (thin solid,
dashed and dotted line respectively) and the axisymmetric anisotropy parameter $k$ (reduced by unity, thick solid line)
for dwarf A. Right panel: the relative differences between dispersion profiles $(\sigma_{\phi} -
\sigma_{\theta})/\sigma_{\theta}$ for dwarfs A, B and C (thin solid, dashed and dotted line respectively) and
$(\sigma_R - \sigma_z)/\sigma_z$ (thick solid line) for dwarf A.
\label{betakdiff}}
\end{figure}

\begin{figure*}
\begin{center}
    \leavevmode
    \epsfxsize=12.5cm
    \epsfbox[0 5 435 430]{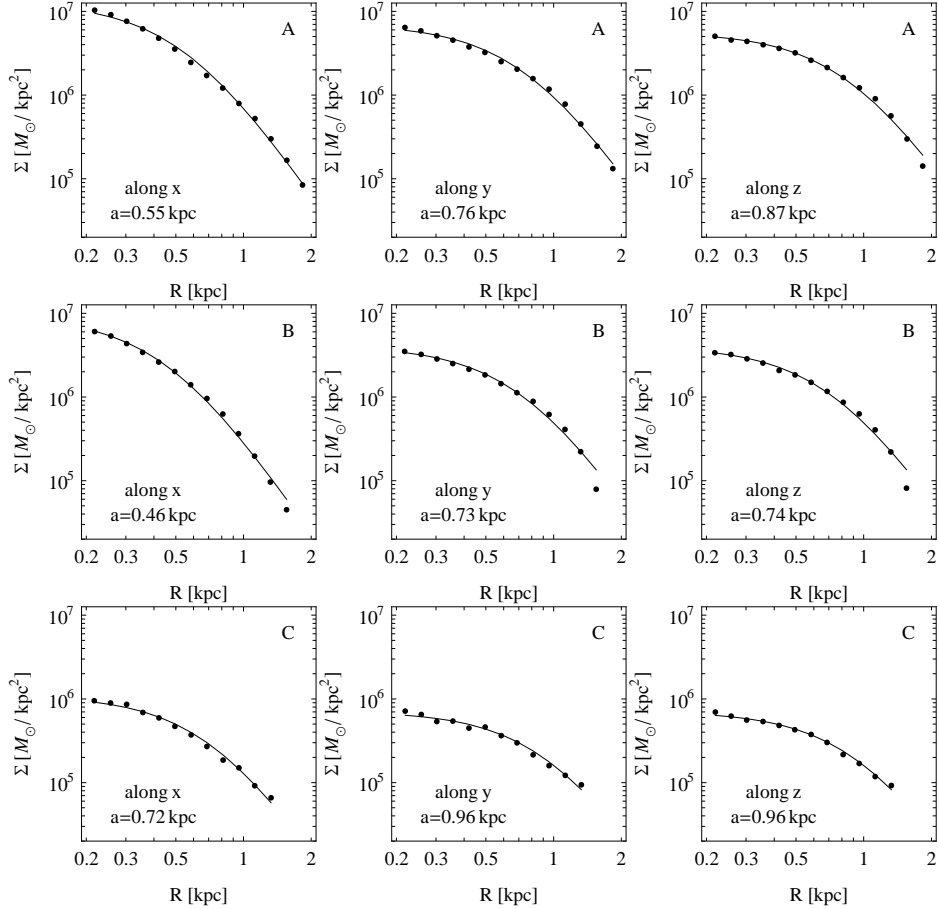}
\end{center}
\caption{Surface density profiles of the stellar component observed along the $x$, $y$ and $z$ axis
(respectively from the left to the right column) for dwarfs formed in simulation A, B and C (from the upper to the
lower row). The filled circles show the measurements made
along the projected radius $R$. The solid lines show
the best fits obtained with the Plummer surface density profile (\ref{projplummer}) with parameters
listed in Table~\ref{fitted} for the spherical case.}
\label{surdenprof}
\end{figure*}

The velocity moments are related to the underlying potential $\Phi$ and the distribution of the tracer $\nu$ via a set
of Jeans equations which in general read (Binney \& Tremaine 2008)
\begin{equation}         \label{jeansgeneral}
     \frac{\partial}{\partial x_i} (\nu \sigma_{ij}^2) + \nu \frac{\partial \overline{v_j}}{\partial t} +
     \nu \overline{v_i} \frac{\partial \overline{v_j}}{\partial x_i} = - \nu \frac{\partial \Phi}{\partial x_j}.
\end{equation}
Together with the continuity equation these equations form a set of four equations relating nine unknown functions, the
three components of the mean velocity $\overline{v_j}$ and six components of the symmetric tensor $\sigma_{ij}^2$. The
system can only be closed if some additional assumptions are adopted, as in the well-studied cases of spherical
symmetry and axisymmetry, which we discuss below. Another possibility, for which Jeans equations are solvable, is the
case of St\"ackel potentials (van de Ven et al. 2003). We find however that none of the well-known triaxial
potentials of this form (de Zeeuw et al. 1986) reproduces the density distribution of all our dwarfs. In particular,
none of the known models reproduces the rather complicated structure of dwarf A, although the distribution of mass in
dwarfs B and C is well fitted by the perfect prolate spheroid (a variation of the so-called perfect ellipsoid, see de
Zeeuw 1985). We will further explore the similarities of those dwarfs to the perfect ellipsoid models elsewhere
({\L}okas et al., in preparation). In any case, with the kind of kinematic samples available for dSph galaxies at
present or in the near future there is little hope of constraining any models beyond spherical or axisymmetric (Statler
1994a,b; van den Bosch \& van de Ven 2009).

\subsection{Spherical models}

We first model only the velocity dispersion profiles using spherical models. For this purpose we solve the
lowest order Jeans equation for spherical systems
\begin{equation}    \label{jeans}
        \frac{\rm d}{{\rm d} r}  (\rho \sigma_r^2) + \frac{2 \beta}{r} \rho
        \sigma_r^2 = - \rho \frac{{\rm d} \Phi}{{\rm d} r}
\end{equation}
where we will assume that mass follows light so $\nu = \rho$ up to a constant factor. The radial velocity
dispersion $\sigma_r$ is related to the angular dispersions $\sigma_{\theta}$ and $\sigma_{\phi}$ by the anisotropy
parameter $\beta=1-\sigma_{\theta}^2/\sigma_r^2$. For simplicity we will consider only models with $\beta=$const i.e.
given by the distribution function $f(E, L) = f(E) L^{-2 \beta}$. Such models require that there is no rotation and
$\sigma_{\theta}=\sigma_{\phi}$. The actual radial dependence of the dispersions (measured with respect to the mean)
$\sigma_r$, $\sigma_{\theta}$ and $\sigma_{\phi}$ on radius is shown in Figure~\ref{dispcomp} for our dwarfs A, B and
C. The corresponding profiles of $\beta$, calculated by replacing $\sigma_{\theta}^2$ with $(\sigma_{\theta}^2 +
\sigma_{\phi}^2)/2$, are plotted in the left panel of Figure~\ref{betakdiff}. The right panel of the Figure shows the
relative differences $(\sigma_{\phi} - \sigma_{\theta})/\sigma_{\theta}$. We can see that the condition
$\sigma_{\theta}=\sigma_{\phi}$ is best obeyed for model B and although $\beta$ profiles vary slowly with radius, our
simulated dwarfs are not strongly anisotropic. We believe this level of anisotropy should have negligible effect
on our findings.

For comparison with observations the
solutions of equation (\ref{jeans}) have to be projected along the line of sight (see e.g. {\L}okas 2002; {\L}okas et
al. 2005). As demonstrated in Figure~\ref{mml}, our models have mass-to-light density ratio almost constant with radius
so in the modeling we assume that the mass follows the distribution of stars. Note that if the dwarfs were not strongly
affected by tides or were in isolation, more general models with extended dark matter profiles would be more
appropriate (e.g. {\L}okas 2002; Koch et al. 2007a; Battaglia et al. 2008; Walker et al. 2009c). The 3D distribution of
stars can be obtained by deprojection of their surface density distribution. The surface density profiles measured for
the observations along the $x$, $y$ and $z$ axis for models A, B and C are plotted as dots in Figure~\ref{surdenprof}.
We have also shown that the density distribution of stars is well approximated by the Plummer law (\ref{plummerden}).
The projected density distribution then has a very simple form
\begin{equation}        \label{projplummer}
        \Sigma(R) = 2 \int_R^\infty \frac{\rho(r) \; r}{\sqrt{r^2-R^2}} {\rm d}r = \frac{a^2 M}{\pi (a^2 + R^2)^2}.
\end{equation}
The fits of this formula to the surface density distributions measured for the three lines of sight are
plotted in Figure~\ref{surdenprof} as solid lines and the fitted scale lengths $a$ are given in each panel of the Figure
and in the first nine rows of Table~\ref{fitted}. (We actually adjusted 2 parameters, the scale length and the
normalization, e.g. the stellar mass, but the latter is not relevant here although it was used in the determination of
the half-light radii listed in Table~\ref{obsprop}. The fits were done with the same weighing scheme as before, only
now applied to surface densities.)

\begin{table}
\caption{Fitted parameters of the surface density profiles. }
\label{fitted}
\begin{center}
\begin{tabular}{ccccc}
\hline
\hline
simulation & model & line of sight & $a$[kpc] & $d$[kpc] \\
\hline
A & spherical    & $x$ & 0.55 &  $-$   \\
  &              & $y$ & 0.76 &  $-$   \\
  &              & $z$ & 0.87 &  $-$   \\
B & spherical    & $x$ & 0.46 &  $-$   \\
  &              & $y$ & 0.73 &  $-$   \\
  &              & $z$ & 0.74 &  $-$   \\
C & spherical    & $x$ & 0.72 &  $-$   \\
  &              & $y$ & 0.96 &  $-$   \\
  &              & $z$ & 0.96 &  $-$   \\
A & axisymmetric & $x$ & 0.56 &  0.77   \\
  &              & $y$ & 0.89 &  1.89   \\
\hline
\end{tabular}
\end{center}
\end{table}

Table~\ref{fittedsph} summarizes the results of fitting the solutions of the Jeans equation (\ref{jeans}) (by the
standard $\chi^2$ minimization) to the velocity dispersion profiles seen along the $x$, $y$ and $z$ axis for models A, B
and C. In each row of the Table the values of the parameters that are kept fixed are given in boldface. We adopted
three approaches. First, assuming the shape of the mass profile given by the fitted scale lengths $a$ we adjusted the
solutions of the Jeans equation for the isotropic case (with the anisotropy parameter $\beta=0$) i.e. fitting only the
total mass $M$. The corresponding solutions are plotted as dotted lines in Figure~\ref{dispprof}. The sixth column of
Table~\ref{fittedsph} gives the masses within $r<r_{\rm max}$ divided by the actual mass of our simulated dwarfs within
this radius. The last column lists the values of the goodness-of-fit measure $\chi^2/N$ rescaled to values
corresponding to the total sample of a thousand stars and a hundred stars per bin. Next, keeping the $a$ value fixed we
adjust $M$ and $\beta$ as free parameters. The corresponding solutions are plotted in Figure~\ref{dispprof} as solid
lines. A third option is to keep $\beta=0$ and adjust $M$ and the scale length $a$. The resulting solutions are plotted
in Figure~\ref{dispprof} as dashed lines. We can see that in most cases by including one more parameter beside the mass
we can significantly improve the quality of the fits. Note that $\beta$ and $a$ cannot both be constrained because they
are strongly degenerate with each other, but any of these parameters improves the fits since they both control the
shape of the dispersion profiles while the mass acts as a normalization.

\begin{table*}
\caption{Fitted parameters of the spherical models.}
\label{fittedsph}
\begin{center}
\begin{tabular}{ccccccc}
\hline
\hline
 simulation & line of sight & $a$ [kpc] & $\beta$ & $M[10^7 {\rm M}_{\odot}]$ &
$\frac{M_{\rm fit}}{M_{\rm true}}(r_{\rm max})$ & $\chi^2/N$ \\
\hline
A &  $x$ & {\bf 0.55} & {\bf 0} & 3.80 & 0.91 & 0.90 \\
  &      & {\bf 0.55} & 0.0020  & 3.80 & 0.91 & 1.02 \\
  &      & 0.70       & {\bf 0} & 4.41 & 0.99 & 0.84 \\
  &  $y$ & {\bf 0.76} & {\bf 0} & 3.25 & 0.71 & 0.45 \\
  &      & {\bf 0.76} & -0.051  & 3.25 & 0.71 & 0.48 \\
  &      & 0.96       & {\bf 0} & 3.76 & 0.74 & 0.35 \\
  &  $z$ & {\bf 0.87} & {\bf 0} & 3.19 & 0.66 & 0.41 \\
  &      & {\bf 0.87} & 0.20    & 3.26 & 0.67 & 0.05 \\
  &      & 0.61       & {\bf 0} & 2.67 & 0.63 & 0.03 \\
\\
B &  $x$ & {\bf 0.46} & {\bf 0} & 2.35 & 1.03 & 0.94 \\
  &      & {\bf 0.46} & 0.26    & 2.45 & 1.07 & 0.34 \\
  &      & 0.29       & {\bf 0} & 1.93 & 0.90 & 0.15 \\
  &  $y$ & {\bf 0.73} & {\bf 0} & 2.11 & 0.80 & 0.63 \\
  &      & {\bf 0.73} & 0.22    & 2.16 & 0.82 & 0.13 \\
  &      & 0.48       & {\bf 0} & 1.73 & 0.75 & 0.07 \\
  &  $z$ & {\bf 0.74} & {\bf 0} & 2.12 & 0.80 & 1.12 \\
  &      & {\bf 0.74} & 0.30    & 2.20 & 0.83 & 0.20 \\
  &      & 0.42       & {\bf 0} & 1.63 & 0.73 & 0.09 \\
\\
C &  $x$ & {\bf 0.72} & {\bf 0} & 1.02 & 1.56 & 1.27 \\
  &      & {\bf 0.72} & -0.20   & 1.04 & 1.59 & 1.10 \\
  &      & 1.65       & {\bf 0} & 1.97 & 1.20 & 0.40 \\
  &  $y$ & {\bf 0.96} & {\bf 0} & 0.92 & 1.13 & 0.40 \\
  &      & {\bf 0.96} & -0.078  & 0.92 & 1.13 & 0.39 \\
  &      & 1.43       & {\bf 0} & 1.23 & 0.94 & 0.24 \\
  &  $z$ & {\bf 0.96} & {\bf 0} & 0.48 & 0.59 & 0.85 \\
  &      & {\bf 0.96} & 0.24    & 0.47 & 0.58 & 0.31 \\
  &      & 0.59       & {\bf 0} & 0.36 & 0.61 & 0.34 \\
\hline
\end{tabular}
\end{center}
\tablecomments{The goodness-of-fit measures $\chi^2/N$ are rescaled to values corresponding to a total sample of a
thousand stars with a hundred stars per bin. The fixed parameters are given in boldface.}
\end{table*}

The mass of dwarf A is underestimated in all approaches, the least so when the line of sight is along the longest axis.
This is obviously due to neglecting the rotation and fitting only the dispersion profiles. This problem can only be
self-consistently solved by considering axisymmetric models with intrinsic rotation, as we do in the next subsection.
The best-fitting anisotropy is zero to mildly radial in rough agreement with the average value measured in 3D for this
dwarf $\beta=0.22 \pm 0.16$ where the uncertainty is the dispersion of $\beta$ values in the left panel of
Figure~\ref{betakdiff}. For dwarf B the mass is typically overestimated when the dwarf is observed along the longest
axis and underestimated if seen along any other axis. Note that only in this case the fitted anisotropy is mildly
radial for all lines of sight and in good agreement with the 3D value of $\beta=0.33 \pm 0.14$. This is probably due to
the fact that for this case the condition $\sigma_{\theta}=\sigma_{\phi}$ required by the spherical models is best
fulfilled (see Figure~\ref{dispcomp} and the right panel of Figure~\ref{betakdiff}). For dwarf C the mass is strongly
overestimated when the dwarf is observed along the longest axis and underestimated along the shortest axis. The
inferred anisotropy varies from mildly tangential to mildly radial, while the 3D value is $\beta=-0.08 \pm 0.17$. In
this case the estimates are least reliable as expected from the highly perturbed state of this dwarf. Note that the
best-fitting values of $a$ differ significantly from the ones obtained by fitting the surface density profiles of the
stars (Table~\ref{fitted}). This approach would therefore only be justified if the density profiles were unknown or
highly uncertain. However, the trends in mass estimates are similar to the case when $a$ are fixed and anisotropy is
fitted instead. With just three cases analyzed, out of which only one (dwarf B) is really reliable (dwarf A has
significant rotation and dwarf C is close to disruption) we cannot draw any firm conclusions concerning the reliability
of our estimates of $\beta$. However, the trend of mass estimates increasing with line of sight closer to the major
axis is clear in all cases.

\subsection{Axisymmetric models}

\begin{figure}
\begin{center}
    \leavevmode
    \epsfxsize=8cm
    \epsfbox[0 0 300 300]{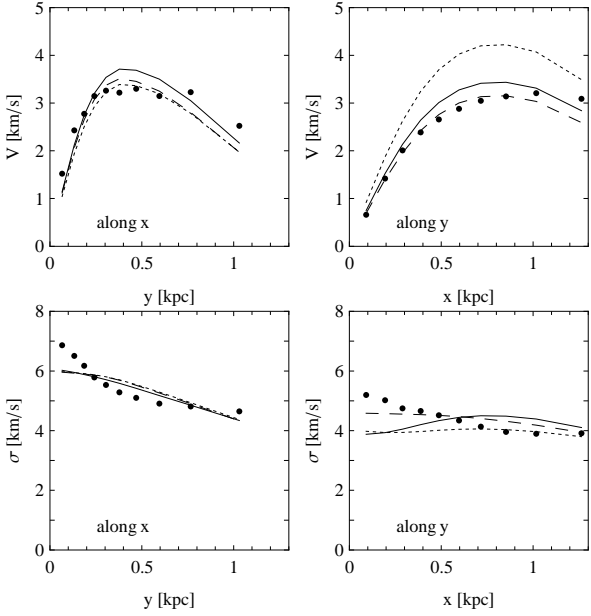}
\end{center}
\caption{The rotation curves (upper row) and the line-of-sight velocity dispersion profiles
(lower row) for the observation along the $x$ and $y$ axis (left and right column respectively) for simulated dwarf A.
The dotted lines show the best fitting solutions of the single component isotropic ($k=1$) models with
scale parameters $a$ and $d$ adopted from the surface density distribution of stars
and only mass $M$ fitted. The dashed lines correspond to best fitting models with $a$,
$d$ and $M$ fitted. The solid lines are for the fits with $M$ and $k$ as free parameters. The best-fitting parameters
are listed in Table~\ref{fittedaxi}.}
\label{veldisp}
\end{figure}

So far we have modeled only the line-of-sight velocity dispersion profiles assuming that the galaxy is spherical.
We will now try to reproduce both the rotation velocity profile and the velocity dispersion for our dwarf A and
those lines of sight where the rotation was detected (observation along the $x$ and $y$ axis).
Both moments are shown as a function of the distance along the major axis of the galaxy image in Figure~\ref{veldisp}.
For the fitting the mean velocities were assigned standard errors of the mean $\sigma/\sqrt{n}$.
For the purpose of this study we will assume that the galaxy can be approximated as an axisymmetric system and again
that the mass distribution follows that of stars. Given that the flattening in our simulated dwarfs is much more
pronounced in the stellar component (see Figure~\ref{density} and fig. 13 in Klimentowski et al. 2009a) than in the
dark halo (which remains almost spherical: the densities measured along the major and the minor axis never differ by
more than a factor of two) it does not seem to be a realistic assumption. However, such one-component models may still
be able to reproduce data better than spherical models.

\begin{figure}
\begin{center}
    \leavevmode
    \epsfxsize=8.5cm
    \epsfbox[0 20 305 165]{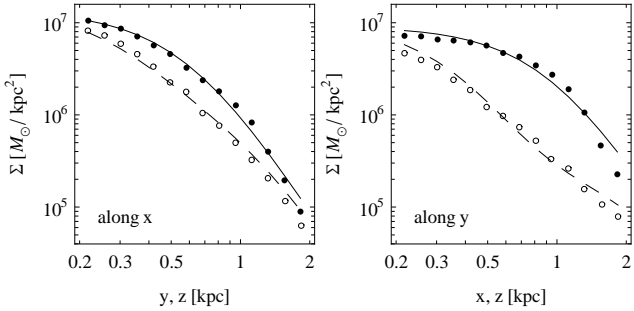}
\end{center}
\caption{Surface density profiles of the stellar component of the dwarf in simulation A seen along the $x$ and $y$ axis
(the left and the right panel, respectively). In both panels
the filled circles show the measurements made in a strip of width $\pm 0.2$ kpc along the major axis of the image
and the open circles the similar measurements along the minor axis. The solid and dashed lines show
the best fits obtained with the projected density associated with potential (\ref{modplum}) with parameters
listed in Table~\ref{fitted} for the axisymmetric case.}
\label{surdenaxi}
\end{figure}

For the axisymmetric systems it is convenient to work in cylindrical coordinates
($R, \phi, z$) where $R$ is now the radial coordinate measured from the center of the galaxy in the equatorial plane.
We will assume that the only streaming motion in the galaxy is the rotation so $\overline{v_R} =
\overline{v_z} =0$, $\overline{v_\phi} \neq 0$. If we further assume that the system is described by a
distribution function that depends only on the energy and the $z$-component of the angular momentum vector, $f=f(E,
L_z)$ then the second velocity moments obey $\overline{v_R^2} = \overline{v_z^2}$, the mixed moments vanish and the
Jeans equations take the form (Binney \& Tremaine 2008)
\begin{eqnarray}
        \frac{\partial}{\partial R} (\rho \overline{v_R^2})
        + \frac{1}{R} \rho (\overline{v_R^2} - \overline{v_{\phi}^2}) &=&
        - \rho \frac{\partial}{\partial R} \Phi
        \label{jeans1}  \\
        \frac{\partial}{\partial z} ( \rho \overline{v_z^2}) &=&
        - \rho \frac{\partial}{\partial z} \Phi.
        \label{jeans2}
\end{eqnarray}

The rotation may be introduced in an arbitrary way, for example we may have (Satoh 1980; Binney et al. 1990)
\begin{equation}         \label{aniso}
        \overline{v_\phi}^2 = k^2 (\overline{v_\phi^2} - \overline{v_R^2})
\end{equation}
where $k$ is the anisotropy parameter and $k=1$ corresponds to the velocity distribution which is isotropic everywhere
in the system, namely $\overline{v_R^2} = \overline{v_z^2} =\overline{v_\phi^2} - \overline{v_\phi}^2$. With this
assumption and the known potential and density, equation (\ref{jeans2}) can be solved for $\overline{v_z^2} =
\overline{v_R^2}$ and introduced into equation (\ref{jeans1}) to obtain the rotation $\overline{v_\phi}$. For
comparison with observations all quantities have to be projected along the line of sight (see Satoh 1980).

The upper right
panel of Figure~\ref{dispcomp} shows that our simulated dwarf is not strongly anisotropic. In this Figure we plot the
profiles of the three velocity dispersions measured in cylindrical shells along $R$ containing equal numbers of
stellar particles. We can see that they are very similar. The only significant difference is visible at $R<1$ kpc where
$\sigma_R$ is somewhat larger than the other two profiles. This is due to the radial motion of the stars in the inner
remnant bar. The relative difference between the dispersions of velocities along $R$ and $z$ is further illustrated in
the right panel of Figure~\ref{betakdiff} as the thick solid line. Its positive value means that the underlying condition
of our axisymmetric models, $\overline{v_R^2} = \overline{v_z^2}$ is not strictly obeyed. The profile of the anisotropy
parameter $k$ (reduced by unity), calculated by replacing $\overline{v_R^2}$ with $(\overline{v_R^2} +
\overline{v_z^2})/2$ in equation (\ref{aniso}), is plotted as the thick solid line in the left panel of
Figure~\ref{betakdiff}.

The density distribution in these axisymmetric systems will be described by the modified Plummer model (\ref{modplum})
with $b=c=0$
\begin{equation}         \label{modplumaxi}
        \Phi(R,z) = -\frac{GM}{[(R^2 + z^2 + a^2)^2 + d^2 z^2]^{1/4}}
\end{equation}
where $R^2=x^2+y^2$, so the model will be fully determined by its total mass $M$ and two scale lengths, $a$ and $d$.
The ratio of the scale lengths $d/a$ controls the amount of flattening and rotation in the system.
We will again start with isotropic ($k=1$) models where
the only free parameter is the mass, while the scale-lengths are determined from the surface density distribution
of the stars. To get them we measure the surface density but this time separately along the major and minor
axis of the galaxy image. The results for the two lines of sight are illustrated in Figure~\ref{surdenaxi}. These
profiles were fitted with the projections of the density following from (\ref{modplumaxi})
and the resulting scale-lengths $a$ and $d$ are listed in the two lower rows of Table~\ref{fitted}.
The quality of the fit is much better for the observation along $x$ (left panel of Figure~\ref{surdenaxi}) than
for the observation along $y$. This is not surprising since in the latter case the bar is more pronounced.

\begin{table*}
\caption{Fitted parameters of the axisymmetric models for simulation A.}
\label{fittedaxi}
\begin{center}
\begin{tabular}{cccccccc}
\hline
\hline
 line of sight & $i$[deg]& $a$ [kpc] & $d$ [kpc]& $k$ & $M[10^7 {\rm M}_{\odot}]$ & $\frac{M_{\rm fit}}{M_{\rm
true}}(r_{\rm max})$ & $\chi^2/N$ \\

\hline
x  &   {\bf 0} & {\bf 0.56} & {\bf 0.77} & {\bf 1} & 4.42 & 1.03 & 0.74 \\
   &   {\bf 0} & {\bf 0.56} & {\bf 0.77} & 1.08    & 4.52 & 1.05 & 0.63 \\
   &   {\bf 0} & 0.54       & 0.76       & {\bf 1} & 4.31 & 1.01 & 0.81 \\

   &  {\bf 10} & {\bf 0.56} & {\bf 0.77} & {\bf 1} & 4.43 & 1.03 & 0.76 \\
   &  {\bf 10} & {\bf 0.56} & {\bf 0.77} &  1.09   & 4.54 & 1.06 & 0.61 \\
   &  {\bf 10} & 0.54       & 0.77       & {\bf 1} & 4.31 & 1.01 & 0.81 \\

   &  {\bf 20} & {\bf 0.56} & {\bf 0.77} & {\bf 1} & 4.46 & 1.04 & 0.82 \\
   &  {\bf 20} & {\bf 0.56} & {\bf 0.77} & 1.12    & 4.60 & 1.07 & 0.56 \\
   &  {\bf 20} & 0.54       & 0.79       & {\bf 1} & 4.35 & 1.02 & 0.82 \\

   &  {\bf 30} & {\bf 0.56} & {\bf 0.77} & {\bf 1} & 4.50 & 1.05 & 1.00 \\
   &  {\bf 30} & {\bf 0.56} & {\bf 0.77} & 1.18    & 4.70 & 1.09 & 0.49 \\
   &  {\bf 30} & 0.53       & 0.84       & {\bf 1} & 4.44 & 1.04 & 0.84 \\

   &  {\bf 40} & {\bf 0.56} & {\bf 0.77} & {\bf 1} & 4.57 & 1.06 & 1.45 \\
   &  {\bf 40} & {\bf 0.56} & {\bf 0.77} & 1.28    & 4.85 & 1.13 & 0.40 \\
   &  {\bf 40} & 0.53       & 0.90       & {\bf 1} & 4.60 & 1.07 & 0.90 \\

   &  {\bf 50} & {\bf 0.56} & {\bf 0.77} & {\bf 1} & 4.62 & 1.08 & 2.39 \\
   &  {\bf 50} & {\bf 0.56} & {\bf 0.77} & 1.46    & 5.04 & 1.17 & 0.32 \\
   &  {\bf 50} & 0.52       & 0.98       & {\bf 1} & 4.85 & 1.13 & 1.07 \\

\\
y  &   {\bf 0} & {\bf 0.89} & {\bf 1.89} & {\bf 1} & 4.38 & 0.81 & 3.85 \\
   &   {\bf 0} & {\bf 0.89} & {\bf 1.89} & 0.84    & 4.11 & 0.76 & 2.28 \\
   &   {\bf 0} & 0.95       & 1.39       & {\bf 1} & 4.58 & 0.85 & 0.62 \\

   &  {\bf 10} & {\bf 0.89} & {\bf 1.89} & {\bf 1} & 4.42 & 0.81 & 3.76 \\
   &  {\bf 10} & {\bf 0.89} & {\bf 1.89} & 0.84    & 4.14 & 0.76 & 2.25 \\
   &  {\bf 10} & 0.95       & 1.40       & {\bf 1} & 4.60 & 0.86 & 0.63 \\

   &  {\bf 20} & {\bf 0.89} & {\bf 1.89} & {\bf 1} & 4.52 & 0.83 & 3.44 \\
   &  {\bf 20} & {\bf 0.89} & {\bf 1.89} & 0.85    & 4.25 & 0.78 & 2.15 \\
   &  {\bf 20} & 0.95       & 1.45       & {\bf 1} & 4.65 & 0.86 & 0.64 \\

   &  {\bf 30} & {\bf 0.89} & {\bf 1.89} & {\bf 1} & 4.70 & 0.87 & 2.84 \\
   &  {\bf 30} & {\bf 0.89} & {\bf 1.89} & 0.87    & 4.46 & 0.82 & 1.95 \\
   &  {\bf 30} & 0.95       & 1.53       & {\bf 1} & 4.78 & 0.88 & 0.68 \\

   &  {\bf 40} & {\bf 0.89} & {\bf 1.89} & {\bf 1} & 4.98 & 0.92 & 1.98 \\
   &  {\bf 40} & {\bf 0.89} & {\bf 1.89} & 0.91    & 4.81 & 0.89 & 1.65 \\
   &  {\bf 40} & 0.95       & 1.67       & {\bf 1} & 5.02 & 0.91 & 0.76 \\

   &  {\bf 50} & {\bf 0.89} & {\bf 1.89} & {\bf 1} & 5.36 & 0.99 & 1.20 \\
   &  {\bf 50} & {\bf 0.89} & {\bf 1.89} & 0.99    & 5.34 & 0.98 & 1.26 \\
   &  {\bf 50} & 0.95       & 1.85       & {\bf 1} & 5.45 & 0.98 & 0.96 \\

\hline
\end{tabular}
\end{center}
\tablecomments{The goodness-of-fit measures $\chi^2/N$ are rescaled to values corresponding to a total sample of a
thousand stars with a hundred stars per bin. The fixed parameters are given in boldface.}
\end{table*}

We proceed to solve the Jeans equations for an axisymmetric system and fit the solutions
to the rotation curves and velocity dispersion profiles measured along the $x$ and $y$ axis and plotted in
Figure~\ref{veldisp}. Table~\ref{fittedaxi} summarizes the results. As for the spherical models, the fixed values of
the parameters are given in boldface and the $\chi^2/N$ values are rescaled to correspond to realistic samples. The
Table lists values obtained for different assumed inclinations of the galaxy where $i$ is the angle between the
rotation plane and the line of sight. The best-fitting profiles of the velocity moments for $i=0^\circ$ are plotted in
Figure~\ref{veldisp}. Again, we tried three approaches, starting with fitting the mass only with $a$ and $d$ adopted
from the fit to the surface density profile and $k=1$. The best-fitting solutions are plotted in Figure~\ref{veldisp}
as dotted lines.

The fits are
quite good in the case of the observation along the $x$ axis but rather poor in the case of observation along the $y$
axis, especially for the rotation curve. The rotation is much too
large compared to the data in spite of fitting the mass. This overshooting of rotation is due to a very large $d/a$
ratio coming from the fit of the surface density distribution of the stars. In other words, self-consistent axisymmetric
and isotropic models would predict more rotation for these $d/a$ ratios. Still, the masses within $r_{\rm max}$ found
in this case are reasonable (see Table~\ref{fittedaxi}), i.e. they are underestimated by no more than 20 per cent.

As for spherical models we next relax the assumption of isotropy and fit the data again with two free parameters,
$M$ and $k$. As discussed in detail by Binney et al. (1990), decreasing $k$ can suppress the rotation curve of the
models while increasing the dispersion profile. The results of the fitting are plotted in Figure~\ref{veldisp} as
solid lines. As expected, the quality of the fits is now improved, especially for the observation along the $y$ axis
(right hand panels of Figure~\ref{veldisp}) where the rotation curve is now in better agreement with the data. Note
however, that the velocity dispersion profile is now worse reproduced and also the masses within $r_{\rm max}$ are even
more underestimated in comparison with the true values. A better approach in this case may therefore be to keep $k=1$
and adjust $a$ and $d$ instead together with the mass. The corresponding profiles of the velocity moments are plotted
in Figure~\ref{veldisp} as dashed lines. The fits are now considerably improved and the masses are not underestimated
by more than 15 per cent. Note also that the fitted values of $a$ and $d$ are not very different from those estimated
from fitting the surface density distribution of the stars.

\begin{figure}
\begin{center}
    \leavevmode
    \epsfxsize=8.1cm
    \epsfbox[0 10 285 420]{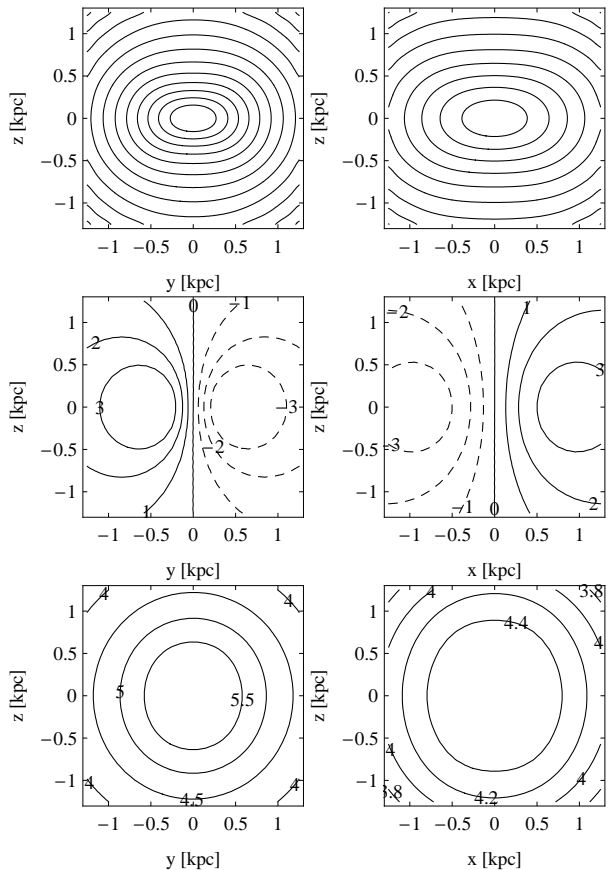}
\end{center}
\caption{The structure and kinematics of the best fitting axisymmetric models for the observation
along the $x$ (left column) and $y$ axis (right column) with fitted parameters $a$, $d$ and $M$ listed
in Table~\ref{fittedaxi} assuming inclination $i=0^\circ$. The upper row panels show the contours of the surface
density $\Sigma$ in M$_{\odot}$ kpc$^{-2}$ plotted at $\Delta \log \Sigma =0.2$ with the innermost contours
corresponding to $\log \Sigma=7.6$ (left column) and 7.2 (right column). The middle and lower row panels show the
contours of the line-of-sight velocity and velocity dispersion respectively, with contour levels in km s$^{-1}$
indicated in each panel. Negative velocity contours are plotted with dashed lines.}
\label{models}
\end{figure}

Figure~\ref{models} shows the surface density contours and the line-of-sight kinematics of these best-fitting
axisymmetric models in a way analogous to Figures~\ref{surden90} and \ref{velocity}. We can see that the models
qualitatively reproduce the shapes and kinematics of the simulated dwarf A. In particular, they reproduce the faster
variation of the velocity moments with projected distance when the dwarf is seen along the longest axis, compared to
the perpendicular direction. Obviously, they are unable to reproduce the velocity gradient seen along the
projected minor axis in the middle panel of the upper row in Figure~\ref{velocity} since this would only
be possible with triaxial models (Binney 1985).

\vspace{0.1in}

\section{Summary and discussion}

Using high-resolution $N$-body simulations we have studied the stellar properties of dSph galaxies resulting from
the tidally induced morphological transformation of disky dwarfs on a cosmologically motivated eccentric orbit around
the Milky Way. We find that although the simulations differed only by the initial inclinations of the disk with respect
to the orbital plane (0$^\circ$,
45$^\circ$, and 90$^\circ$), this single parameter change led to significantly dissimilar
outcomes. The reason for this is the well-known effect that the tidal force is more effective
in removing stars which are on more prolate orbits with respect to the orbital motion (e.g., Read et al.
2006). The dwarf spheroidal galaxies obtained had different masses, density distributions,
and velocity structure. Although we have used a most probable orbit of dwarf galaxies in the Local Group (Diemand et
al. 2007), the variation of orbital parameters would further differentiate the outcome of the simulations. On more
circular orbits the dwarfs would be less affected by impulsive heating at pericenters and therefore we expect them to
be less evolved, preserving a more prolate shape and more rotation as in earlier stages of the evolution on our
eccentric orbit (see Klimentowski et al. 2009a).

The least evolved dwarf A (with initial inclination of $90^\circ$) has an intrinsically triaxial distribution in the
stellar component and preserved a fair amount of rotation. The rotation is mostly around the shortest axis. The
intermediate case, dwarf B (with initial inclination of $45^\circ$), is a prolate spheroid and retained
little rotation. The shape of its stellar component is still rather regular and so the dwarf can be considered a
relaxed object. The most strongly evolved dwarf C (with initial inclination of $0^\circ$) is also a prolate spheroid
with very little rotation but its noisy density contours and profiles of velocity moments indicative of small numbers
of bound stars suggest that it is very close to disruption.

Despite the aforementioned differences and the fact that the final masses of the simulated dwarfs differ
significantly (see Table~\ref{simprop} and Figure~\ref{mml}), the final $M/L$ ratios are quite similar and
fairly constant with radius. The former is related to the fact that the initial mass-to-light distributions
of the dwarfs were identical and indicates that the dark matter and stars are lost in proportional
amounts in all cases. The latter is a consequence of both our choice of initial conditions for the dwarf
galaxy models and the fact that most of the dark matter outside of the stellar component is stripped efficiently.
Indeed, in the present study we adopt the Mo et al. (1998) scalings which connect the angular
momentum/size of the disk with those of the parent dark matter halo. These general scalings together with
our specific choice of parameters result in an initial dwarf galaxy with a $M/L$ ratio which is constant
within $2$ kpc or so and increases thereafter (Klimentowski et al. 2007). As the outer dark matter halo
gets tidally stripped, the effect of increasing $M/L$ ratio at larger distances, characteristic of the initial
conditions, disappears. However, it is not difficult to imagine scenarios where this situation can change as,
for example, in cases where the original disky dwarfs are constructed with much smaller scale lengths compared
to the scale radius of the parent halo, $r_s$. Such modeling would probably result in a non-constant $M/L$ ratio
within the luminous radius of the dwarf as the outer dark matter halo is tidally truncated. Until cosmological simulations
reveal the structure of the dwarf progenitors themselves, the choice of initial structural parameters will remain
somewhat arbitrary.

We studied the density distribution of the stellar component both radially averaged and along the principal axes. In
the former case it is well fitted by the Plummer law, in agreement with real dSph galaxies. For the latter we proposed
a simple formula for the potential of the stellar component of the simulated dwarfs based on the generalization of the
Plummer law. This formula reproduces quite well the triaxial shape of model A and prolate spheroidal shapes of models B
and C (with a smaller number of parameters in the latter cases) both when measured along the principal axes and in
projection. The surface density distributions are quite similar to the observed ones, although a detailed comparison is
beyond the scope of the present paper.

We also investigated the velocity distribution of the stars in the dwarfs along the principal axes and along the line of
sight. The velocity structure seems to be the richest in the case of model A due to its triaxiality. In this case we
detect rotation both around the short and the long axis which manifests itself in the misalignment of the zero velocity
curve with respect to the projected minor axis. The velocity structure is much simpler in dwarfs B and C since almost
no net streaming motion is seen there. In all models the dispersion of velocity is largest for velocities along the
major axis which is due to the origin of the shapes of the dwarfs. All of them went through a prolonged
bar-like phase where radial orbits along the bar are dominant and were transformed into prolate spheroids by systematic
shortening of the bar.

The projected velocity dispersions of our simulated galaxies decline slowly with radius,
although some exhibit larger gradients. This behavior agrees with the observed profiles
in real dSph galaxies (e.g. {\L}okas et al. 2008; {\L}okas 2009) when the samples are cleaned of contamination
using algorithms such as the one described in Klimentowski et al. (2007). Note that the majority of velocity
dispersion profiles published in the literature (e.g. Mu\~noz et al. 2005; Koch et al. 2007b; Walker et al.
2007) are rather flat, which in our opinion is probably due to contamination from the Milky Way or tidally stripped
stars but may also be caused by multiple stellar populations (Battaglia et al. 2006) or an extended dark matter
component (e.g. Walker et al. 2006).

This velocity structure has important consequences for mass modeling. We fitted the line-of-sight velocity dispersion
profiles of the dwarfs observed along the principal axes using different sets of free parameters. When the observation
is done along the longest axis (the velocity dispersion is then larger and declines more steeply with radius) the mass
is typically mildly overestimated. When it is done along the intermediate or short axis, it is underestimated.

The effect is largest for dwarf C, which is close to disruption, but even in this case the mass is over/underestimated by
only about 60 percent. This suggests that the very high mass-to-light ratios estimated for some dSph galaxies, such as
Draco or Ursa Major, cannot be explained by departures from sphericity, but are rather the result of the formation
process of their progenitors, or are caused by other mechanisms that affected their baryonic mass fraction such as
photoevaporation of the gas after reionization (e.g. Babul \& Rees 1992; Bullock et al. 2000) or
the combined effect of heating by the UV background and subsequent ram pressure stripping in the hot gaseous corona
of the Milky Way (Mayer et al. 2006; 2007).

For dwarf B, which seems to provide the best model for the real dSph galaxies (it still has a regular structure and
shows little rotation) we find that spherical modelling may overestimate the mass by up to 7 percent (when the line of
sight is along the major axis) but underestimate it by up to 27 percent (when viewed along a short axis). Large
kinematic samples of a few hundred up to a few thousand stars with measured velocities recently obtained by Walker et
al. (2009b) for Carina, Fornax, Sculptor and Sextans allow for mass determinations with statistical errors as small as
5-15 percent ({\L}okas 2009) using simple models (assuming constant mass-to-light ratios and anisotropy) described here
or similar. This means that the systematic errors due to non-sphericity of dSph galaxies are comparable or larger than
the statistical errors in mass determination due to sampling errors of velocity moments for the best-studied dwarfs.

For dwarf A the velocity moments (the rotation curve and the velocity dispersion profile) are much better reproduced
with axisymmetric models constructed under the assumption that only the rotation around the shortest axis in present
(i.e. neglecting the rotation around the longest axis). Although the presence of rotation has been
reported in a number of dSph galaxies, like Ursa Minor (Hargreaves et al. 1994; Armandroff et al. 1995), Leo I
(Sohn et al. 2007), Sculptor (Battaglia et al. 2008), Carina (Mu\~noz et al. 2006) and recently for more distant
ones, Cetus (Lewis et al. 2007) and Tucana (Fraternali et al. 2009), it remains to be seen if such models are
applicable to real dSph galaxies as they would require measurements of rotation curves at higher significance level
than presently available.

We have demonstrated that the kinematics of our simulated dwarfs depends sensitively
on the line of sight. This has intriguing implications for the missing satellites problem (Klypin et al. 1999;
Moore et al. 1999), which is formulated by mapping the observed stellar velocity dispersions of dwarf galaxies
to subhalo maximum circular velocities, $V_{\rm max}$ (see also Klimentowski et al. 2009a). These results imply that
because our remnant dwarfs are non-spherical, $V_{\rm max}$ depends both on the specific model used to fit the
potential of the dwarf and the line of sight used to measure stellar velocity dispersions, $\sigma$. In our models, the
measurement of the central velocity dispersion alone, when done along the longest axis can be higher by 50 percent
compared to the perpendicular direction. We note that these findings are only relevant to dSphs that have a moderate
mass-to-light ratio like our simulated dwarfs (e.g. Fornax or Leo I). Dwarfs embedded in a much more massive dark
matter halo may in principle have a different relation between $\sigma$ and $V_{\rm max}$ since the stars could probe a
region well inside the radius at which $V_{\rm max}$ occurs.

Lastly, the findings of the present study demonstrate that for a given orbit and initial density distribution
the stellar structure and kinematics of tidally stripped disky dwarfs depend sensitively on the
initial inclination of the disk with respect to the orbital plane. Similar conclusions were reached by
Mastropietro et al. (2005) using $N$-body simulations to follow the tidal evolution and harassment
of disk galaxies inside a galaxy cluster environment. The models presented in this paper as well as those of
Mastropietro et al. (2005) do not include gas.

Mayer et al. (2006, 2007) have shown that gas present in the disky
progenitors of dSphs can be either completely stripped (by ram pressure and tides) or partially retained producing
periodic bursts of star formation at pericentric passages. The outcome depends sensitively on when the dwarf fell into
the primary potential. The discriminating factors are the strength of the cosmic ultraviolet background and the local
flux from the primary at the time of infall. In cases when gas is partially retained, it can affect the structural
properties of the remnants, favoring a longer lived stellar bar and reducing tidal mass loss by increasing its central
density following a bar-driven gas inflow. This would tend to produce a remnant more similar to that of model A of this
paper, i.e. triaxial and with more residual rotation. However, as the gas is consumed and the dwarf becomes devoid of
it, further tidal heating may change the internal structure of the dwarf again. Fornax, Carina and Leo I, that had
prolonged star formation, may have gone through these different phases. Overall, the inclusion of gas is likely to
induce further scatter in the final structural properties of dwarf galaxies orbiting the Milky Way. We will
extend our analysis to gasdynamical models in future work.

\acknowledgments

The numerical simulations were performed on the zBox2 supercomputer at the University of Z\"urich.
SK is funded by the Center for Cosmology and Astro-Particle Physics at The Ohio State University.
This research was partially supported by the Polish Ministry of Science and Higher Education
under grant NN203025333.

%\newpage

\end{document}